\documentclass[english,draftcls, journal, 12pt, one column]{IEEEtran}
\usepackage[T1]{fontenc}
\usepackage[latin9]{inputenc}
\usepackage{verbatim}
\usepackage{bm}
\usepackage{amsmath}
\usepackage{amssymb}
\usepackage{graphicx}

\makeatletter

\providecommand{\tabularnewline}{\\}

\IEEEoverridecommandlockouts

\usepackage{amsfonts}\usepackage{mathrsfs}\usepackage{bm}\usepackage{algorithmicx}
\usepackage{array}\usepackage{mdwmath}\usepackage{mdwtab}\usepackage[caption=false,font=footnotesize]{subfig}\usepackage{fixltx2e}
\usepackage{footnote}\usepackage{tabularx}\usepackage{multirow}\usepackage{cite}
\usepackage{fixltx2e}\usepackage{braket}\usepackage{algorithmicx}\usepackage{algpseudocode}\usepackage{multirow}

\usepackage{url}
\makeatletter

\newcommand{\Rmnum}[1]{\expandafter\@slowromancap\romannumeral #1@}
\makeatother

\makeatother

\usepackage{babel}
\begin{document}

\title{\textbf{\normalsize{}Efficient Interference Management Policies for
Femtocell Networks}}

\author{\IEEEauthorblockN{{\normalsize{}Kartik Ahuja, Yuanzhang Xiao and
Mihaela van der Schaar }}\\
{\normalsize{} \IEEEauthorblockA{Department of Electrical Engineering,
UCLA, Los Angeles, CA, 90095}\\
{\normalsize{} Email: ahujak@ucla.edu, yxiao@seas.ucla.edu and mihaela@ee.ucla.edu}}
\vspace{-2em}\thispagestyle{plain}\vspace{-1.35em} \pagestyle{plain}}

\maketitle
\vspace{-1em}

\begin{abstract}
Managing interference in a network of macrocells underlaid with femtocells
presents an important, yet challenging problem. A majority of spatial
frequency/time reuse based approaches partition the users by coloring
the interference graph, which is shown to be suboptimal. Some spatial
time reuse based approaches schedule the maximal independent sets
(MISs) of users in a cyclic, (weighted) round-robin fashion, which
is inefficient for delay-sensitive applications. Our proposed policies
schedule the MISs in a $\emph{non-cyclic}$ fashion, which aim to
optimize any given network performance criterion for delay-sensitive
applications while fulfilling minimum throughput requirements of the
users. Importantly, we do not take the interference graph as given
as in existing works; we propose an optimal construction of interference
graph. We prove that under certain conditions, the proposed policy
achieves the optimal network performance. For large networks, we propose
a low-complexity algorithm for computing the proposed policy, and
prove that under certain conditions, the policy has a competitive
ratio (with respect to the optimal network performance) that is independent
of the network size. The proposed policy can be implemented in a decentralized
manner by the users. Compared to the existing policies, our proposed
policies can achieve improvement of up to 130 \% in large-scale deployments.


\end{abstract}

\section{Introduction}

As more and more devices are connecting to cellular networks, the
demand for wireless spectrum is exploding. Dealing with this increased
demand is especially difficult because most of the traffic comes from
bandwidth-intensive and delay-sensitive applications such as multimedia
streaming, video surveillance, video conferencing, gaming etc. These
demands make it increasingly challenging for the cellular operators
to provide sufficient quality of service (QoS). Dense deployment of
distributed low-cost femtocells (or small cells in general, such as
microcells and picocells) has been viewed as one of the most promising
solutions for enhancing access to the radio spectrum\cite{ghosh2012heterogeneous}\cite{andrews2012femtocells}.
Femtocells are attractive because they can both extend the service
coverage and boost the network capacity by shortening the access distance
(cell splitting gain) and offloading traffic from the cellular network
(offloading gain). However, in a closed access network when only registered
mobile users can connect to the femtocell base station, dense deployment
of femtocells operating in the same frequency band leads to strong
co-tier interference. In addition,  the macrocell users  usually
operate in the same frequency, the problem of interference (to both
femtocells and macrocells) is further exacerbated due to cross-tier
interference across macrocells and femtocells. In this work, we study
a closed access network. Hence, it is crucial to design interference
management policies to deal with both co-tier and cross-tier interference.

Interference management policies specify the transmission schedule
and transmit power levels of femtocell user equipments FUEs (femtocell
base stations FBSs) and macrocell user equipments MUEs (macrocell
base stations MBSs) in uplink (downlink) transmissions. \footnote{For brevity, we will focus on uplink transmissions hereafter.}
An efficient interference management policy should fulfill the following
important requirements (we will discuss  in Section \ref{sec:related works},
state-of-the-art policies do not fulfill one or more of the following
 requiremenst):
\begin{itemize}
\item ${\it \textit{Interference management based on network topology}}$:
Policies must take into account that uplink transmissions from neighboring
UEs create strong mutual interference, but must also recognize and
take advantage of the fact that non-neighboring UEs do not. Hence,
the network topology (i.e. locations of femtocells/macrocells) must
play a crucial role.
\item $\textit{Limited signaling for interference coordination}$: In dense,
large-scale femtocell deployments, the UEs cannot coordinate their
transmissions by sending a large amount of control signals across
the network. Hence, effective policies should not rely on heavy signaling
and/or message exchanges across the UEs in the network.
\item ${\it \textit{Scalability in large networks}}$: Femtocell networks
are often deployed on a large scale (e.g. in a city). Effective  policies
should scale in large networks, i.e. achieve efficient network performance
while maintaining low computational complexity.

\item ${\it \textit{Support for delay-sensitive applications}}$: Effective
 policies must support delay-sensitive applications, which constitute
the majority of wireless traffic.


\item ${\it \textit{Versatility in optimizing various network performance criteria}}$:
The appropriate network performance criterion (e.g. weighted sum throughput,
max-min fairness, etc.) may be different for different networks. Effective
 policies should be able to optimize a variety of network performance
criteria while ensuring performance guarantees for each MUE and each
FUE.
\end{itemize}
In this work, we propose a novel, systematic, and practical methodology
for designing and implementing interference management policies that
fulfill $\emph{all}$ of the above requirements. Our proposed policies
aim to optimize a given network performance criterion, such as weighted
sum throughput and max-min fairness, subject to each UE's minimum
throughput requirements. Our proposed policies can efficiently manage
a wide range of interference. We manage strong interference between
neighboring UEs by using time-division multiple access (TDMA) among
them. We take advantage of weak interference between non-neighboring
UEs by finding maximal sets of UEs that do not interfere with each
other and allowing all the UEs in those sets to transmit at the same
time. Specifically, we find the maximal independent sets (MISs) \footnote{A set of vertices in which no pair is connected by an edge is independent
(IS) and if it is not a subset of another IS then it is MIS. } of the interference graph, and schedule different MISs to transmit
in different time slots. The scheduling of MISs in our proposed policy
is particularly designed for delay-sensitive applications: the schedule
of MISs across time is not cyclic (i.e. the policies do not allocate
transmission times to MISs in a fixed (weighted) round-robin manner),
but rather follows a carefully designed nonstationary schedule, in
which the MIS to transmit is determined adaptively online. For delay-sensitive
applications, cyclic policies are inefficient because transmission
opportunities (TXOPs) earlier in the cycle are more valuable than
TXOPs later in the cycle (earlier TXOPs enhances the chances of transmission
before delay deadlines). The cyclic polices are unfair to UEs allocated
to later TXOPs.

Another distinctive feature of our work is that we do not take the
interference graph as given as in most existing works; instead, in
our work we show how to choose the interference graph that maximizes
the network performance. Specifically, in our construction of interference
graphs, we determine how to choose the threshold on the distance between
two cells, based on which we determine if there is an edge between
them, in order to maximize the network performance. Moreover, we prove
that under certain conditions, the proposed policy, computed based
on the optimal threshold, can achieve the optimal network performance
(weighted sum throughput) within a desired small gap. Note that for
large networks, in general it is computationally intractable  to find
all the MISs of the interference graph \cite{johnson1988generating}.
We propose efficient polynomial-time algorithms to find a subset of
MISs, and prove that under certain conditions, the proposed policy,
computed based on the constructed subset of MISs, can achieve a constant
competitive ratio (with respect to optimal weighted sum throughput)
that is independent of the network size.

Finally, we summarize the main contributions of our work:

1. We propose interference management policies that schedule the MISs
of the interference graph. The schedule  is constructed in order to
maximize the network performance criterion subject to minimum throughput
requirements of the UEs. Moreover, the schedule adapts to the delay
sensitivity requirements of the UEs by scheduling transmissions in
a non-stationary manner.

2. We construct the interference graph by comparing the distances
between the BSs with a threshold (i.e. there is an edge  between two
cells if the distance between their BSs is smaller than the threshold).
We develop a procedure to choose the optimal threshold such that the
proposed scheduling of MISs leads to a high network performance. Importantly,
we prove that under certain conditions, the proposed scheduling of
MISs based on the optimal threshold achieves within a desired small
gap of the optimal network performance (weighted sum throughput).

3. It is intractable to find all the MISs in large networks, we propose
an approximate algorithm that computes a subset of MISs within polynomial
time. We prove that under certain conditions, the proposed policy
based on this subset of MISs has a constant competitive ratio (with
respect to the optimal weighted sum throughput) that is independent
of the network size.


The rest of the paper is organized as follows. In Section \ref{sec:related works}
we discuss the related works and their limitations followed by the
system model and  problem formulation in Section \ref{sec:sys model}
and \ref{sec:prblm formuln} respectively. The design framework and
its low-complexity variant for large networks are discussed in Section
\ref{sec:design framework} and Section \ref{sec:low complexity implem}
respectively. In Section \ref{sec:simulations} we use simulations
to compare the proposed policy with state-of-the-art  followed by
conclusion in Section \ref{sec:conclusion}.

\section{Related Works}

\label{sec:related works} In this section we compare our proposed
policy with state-of-the-art policies. Existing interference management
policies can be categorized in two classes: 1) policies based on power
control, and 2) policies based on spatial time/frequency reuse.

\vspace{-1em}

\subsection{Interference Management Policies Based on Power Control}

The first and most widely-used interference management policies\cite{Decen-pow-game,dist-pow-ofdma,dist-res-femto,giupponi2010distributed,hv-zhu,jointerf,andrews-1}
are based on power control. In these policies, all the UEs in the
network transmit at a $\emph{constant}$ power at all time (provided
that the system parameters remain the same) in the entire frequency
band \footnote{Although some power control policies \cite{Decen-pow-game} go through
a transient period of adjusting the power levels before converging
to the optimal power levels, the users maintain the constant power
levels after the convergence.}. When the cross channel gains among BSs and UEs are high, simultaneous
transmissions at the same time and in the same frequency band will
cause significant interference among cells. Such strong interference
is common in macrocells underlaid with femtocells. For example, in
\cite{choi2008dealing} it is shown that that interference from MUEs
near the FBS severely affects the uplink transmissions of FUEs. Also,
in offices and apartments, where FBSs are installed close to each
other, inter-cell interference is particularly strong. In contrast,
our proposed solutions mitigate the strong interference by letting
only a subset of non-interfering UEs to transmit at the same time.

\subsection{Interference Management Policies Based on Spatial Time/Frequency
Reuse}

Some existing works mitigate strong interference by letting different
subsets of UEs to transmit in different time slots (spatial time reuse)
\cite{ramanathan1993scheduling,huson1995broadcast,ramaswami1989distributed,cidon1989distributed,pateromichelakis2012dynamic,brarcomputationally,ephremides1990scheduling,aggarwal2011achieving,jain2005impact}
or in different frequency channels (spatial frequency reuse) \cite{graph-cluster,graph-dynamic,lee2010interference,lopez2009ofdma,kim2008throughput,saquib2012interference,necker2008graph}.
Specifically, they partition UEs into disjoint subsets such that the
UEs in the same subset do not interfere with each other. Given the
same partition of the UEs, the policies based on spatial time reuse
and those based on spatial frequency reuse are equivalent. Hence,
we focus on policies based on spatial time reuse hereafter.

Some policies based on spatial time reuse, partition the UEs based
on the coloring of the interference graph h \cite{ramanathan1993scheduling,huson1995broadcast,ramaswami1989distributed,cidon1989distributed,pateromichelakis2012dynamic},
which is not efficient. In general, a set of UEs with the same color
(i.e. the UEs who can transmit simultaneously) may not be maximal
(See Fig. 1 a), in the sense that there may be UEs who do not interfere
but have different colors (we will also show this in the motivating
example in Subsection \ref{sub:Motivating-Example}). In this case,
it is more efficient to also let those non-interfering UEs to transmit
simultaneously, although they have different colors. Hence, the partitioning
based on coloring the interference graph is not efficient, because
the average number of active UEs (i.e. the average cardinality of
the subsets of UEs with the same color) is low.

Some policies based on spatial time reuse\cite{brarcomputationally,ephremides1990scheduling,jain2005impact,aggarwal2011achieving}
partition the UEs based on the MISs of the interference graph, which
is more efficient, because we cannot add any more UEs to an MIS without
creating strong interference. However, they are still inefficient
compared to our proposed policies for delay-sensitive applications.
Specifically, they schedule different MISs in a cyclic and (weighted)
round-robin manner, in which each UE transmits at a fixed position
in each cycle. For delay-sensitive applications, earlier positions
in the cycle are more desirable because they enhance the chances of
transmitting prior to delay deadlines. Hence, a cyclic schedule is
not fair to the UEs allocated to later positions. In contrast, our
proposed policies schedule the MISs in an efficient, nonstationary
manner for delay-sensitive applications.

Another notable difference from the existing works based on spatial
time/frequency reuse is that they usually take the interference graph
as given. On the contrary, our work discusses how to construct the
interference graph optimally such that the network performance is
maximized.

\subsection{Other Interference Management Policies}

Besides the above two categories, there are several other related
works. For instance in \cite{nazir2010learning}\cite{bennis2013self},
the authors propose reinforcement learning and evolutionary learning
techniques for the femtocells to learn efficient interference management
policies. In \cite{nazir2010learning}, the femtocells learn the fixed
transmit power levels, while in \cite{bennis2013self}, the femtocells
learn to randomize over different transmit power levels. However,
the interference management policies in \cite{nazir2010learning}
and \cite{bennis2013self} cannot provide minimum throughput guarantees
for the UEs. In constrast, we provide rigorous minimum throughput
guarantees for the UEs. In both \cite{nazir2010learning}\cite{bennis2013self}
the femtocell UEs need to limit their transmission powers in every
time slot such that the SINR of the macrocell UE is sufficiently high.
If there is strong interference between some femtocells and the macrocell,
the femtocell UEs will always transmit at lower power levels, leading
to a low sum throughput for them.

Another method to mitigate interference is to deploy coordinated beam
scheduling \cite{chae2012interference}\cite{yang2011distributed}.
In \cite{chae2012interference} and \cite{yang2011distributed}, the
authors schedule a subset of beams to maximize the total reward associated
with the scheduled subset, where the reward per beam reflects the
channel quality and traffic. The first difference from our work is
that the proposed approach schedules a fixed subset of beams and leaves
the other UEs inactive. Hence, some UEs have no throughput, which
means the minimum throughput as well as the delay-sensitivity of the
UEs is not satisfied. Second, we rigorously prove that our proposed
policy achieves good performance with low (polynomial-time) complexity
, while \cite{chae2012interference}\cite{yang2011distributed} do
not. Third, the schemes in \cite{chae2012interference}\cite{yang2011distributed}
are proposed for a specific network performance criterion and may
not be flexible enough for other network performance criteria (such
as the sum throughput). Finally, \cite{chae2012interference}\cite{yang2011distributed}
do not consider delay sensitivity of the UEs.

\vspace{-0.8em}

\section{System Model}

\label{sec:sys model}
\begin{figure}
\begin{minipage}[t]{0.45\textwidth}%
\includegraphics[width=2.9in,height=1.4in,keepaspectratio]{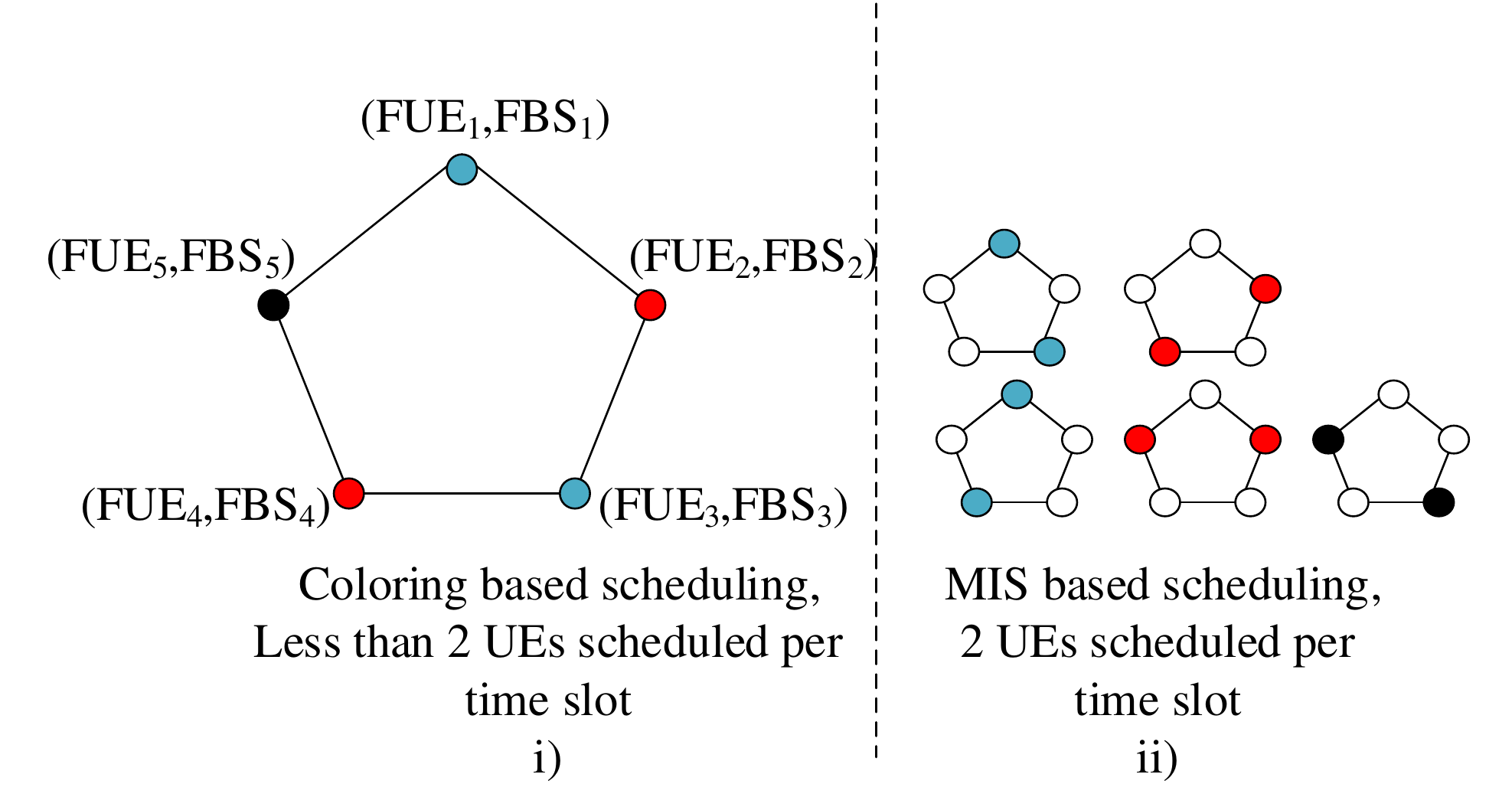}

\tiny\text{Fig. 1 a)}%
\end{minipage}\quad{}%
\begin{minipage}[t]{0.35\textwidth}%
\includegraphics[width=2.9in,height=1.8in]{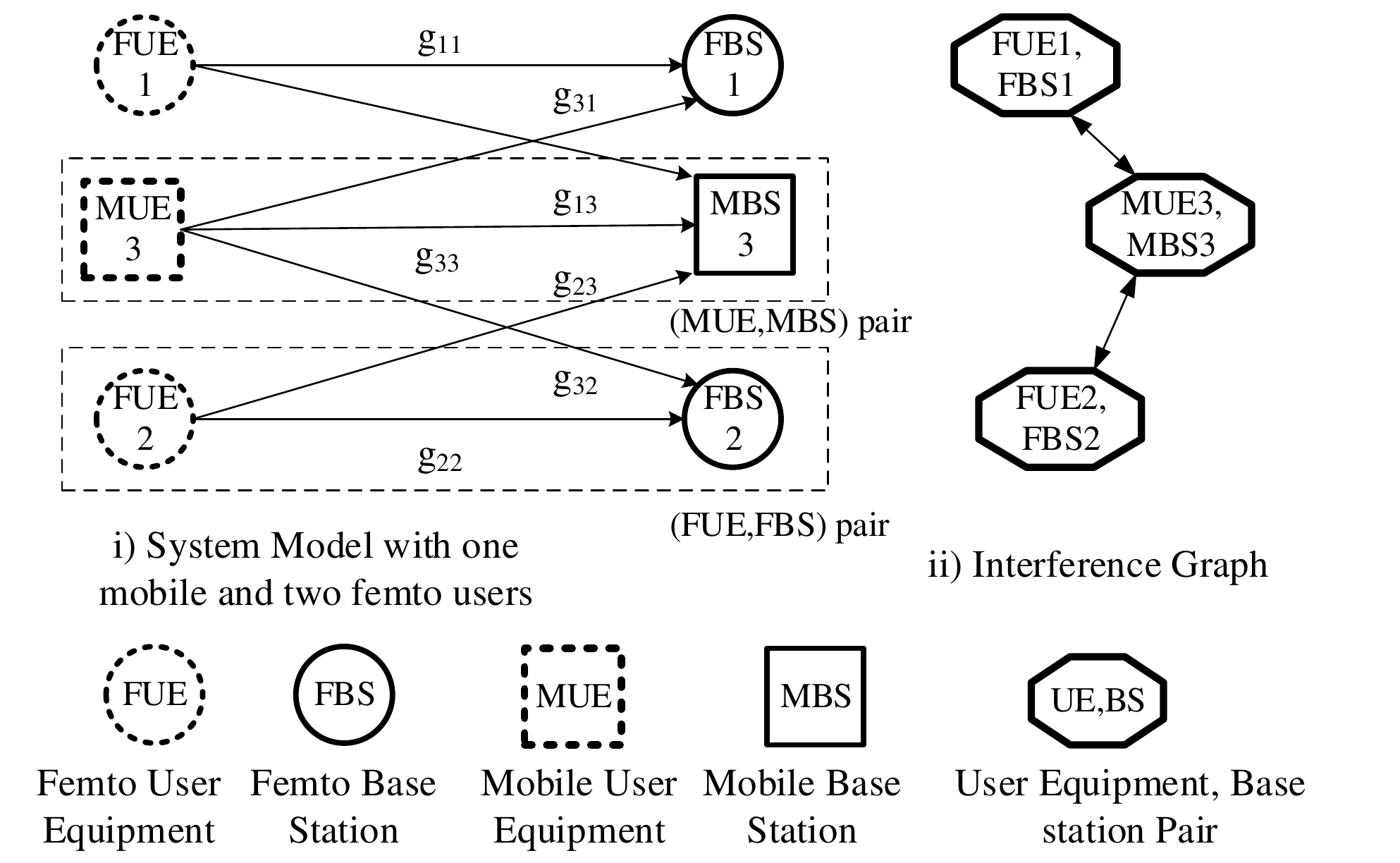}

\tiny\text{Fig. 1 b)}%
\end{minipage} \protect\caption{a) Coloring based scheduling in i) schedules less than two UEs per
time slot on an average, while MIS based scheduling in ii) is more
efficient and schedules two UEs per time slot. , b) System model illustration. }

\end{figure}

\subsection{Heterogeneous Network of Macrocells and Femtocells}

We consider a heterogeneous network of $N$ femtocells (indexed by
$\{1,2,\ldots,N\}$) and $M$ macrocells (indexed by $\{N+1,...,N+M\}$)
operating in the same frequency band, a common deployment scenario
considered in practice \cite{Decen-pow-game}. We assume that each
FBS/MBS serves only one FUE/MUE . Our model can be easily generalized
to the setting where each BS serves multiple UEs, at the expense of
complicated notations to denote the association among UEs and BSs.
For notational clarity, we focus on the case where each BS serves
one UE, and will demonstrate the applicability of our work to the
setting where one BS serves multiple UEs in Section \ref{sec:simulations}.

Since there is only one FUE or MUE in a femtocell or a macrocell,
the index of each UE and that of each BS are the same as the index
of the cell they belong to. We focus on the uplink transmissions.
The proposed framework can be applied directly to the downlink scenarios
in which each BS serves one UE at a time. See Fig. 1b)  for an illustration
of a 3-cell network with $N=2$ femtocells and $M=1$ macrocell.%
{} Each UE $i$ chooses its transmit power $p_{i}$ from a compact set
$\mathcal{P}_{i}\subseteq\mathbb{R}_{+}$. We assume that $0\in\mathcal{P}_{i},\;\forall i\;\in\{1,...N+M\}$,
namely a UE can choose not to transmit. The joint power profile of
all the UEs is denoted by $\textbf{p}=(p_{1},....p_{N+M})\in\mathcal{P},$
where $\mathcal{P}=\prod_{i=1}^{N+M}\mathcal{P}_{i}$. The power profile
of all the UEs other than $i$ is denoted by $\textbf{p}_{-i}$. When
a UE $i$ chooses a transmit power $p_{i}$, the signal to interference
and noise ratio (SINR) experienced at BS $i$ is $\gamma_{i}(\textbf{p})=\frac{g_{ii}p_{i}}{\sum\limits _{j\not=i}g_{ji}p_{j}+\sigma_{i}^{2}}$,
here $g_{ji}$ is the channel gain from UE $j$ to BS $i$, and $\sigma_{i}^{2}$
is the noise power at BS $i$. 
Since the BSs cannot cooperate to decode their messages, each BS $i$
treats the interference as white noise, and gets the following throughput
\cite{Decen-pow-game} $r_{i}(\textbf{p})=\log_{2}(1+\gamma_{i}(\textbf{p}))$.

\subsection{Interference Management Policies}

The system is time slotted at $t$ = 0,1,2..., and the UEs are assumed
to be synchronized as in \cite{etkin2007spectrum}. At the beginning
of time slot $t$, each UE $i$ decides its transmit power $p_{i}^{t}$
and obtains a throughput of $r_{i}(\textbf{p}^{t})$. Each UE $i$'s
strategy, denoted by $\pi_{i}:\mathbb{Z}_{+}=\{0,1,..\}\rightarrow\mathcal{P}_{i}$,
is a mapping from time $t$ to a transmission power level $p_{i}\in\mathcal{P}_{i}$.
The interference management policy is then the collection of all the
UEs' strategies, denoted by $\boldsymbol{\pi}=(\pi_{1},...,\pi_{N+M})$.
Each UE is ${\it {delay\;sensitive}}$ and hence discounts the future
throughputs as in \cite{xiao2012dynamic}. The average discounted
throughput for UE $i$ is given as $R_{i}(\boldsymbol{\pi})=(1-\delta)\sum\limits _{t=0}^{\infty}\delta^{t}r_{i}(\textbf{p}^{t}),$where
$\textbf{p}^{t}=(\pi_{1}(t),...,\pi_{N+M}(t))$ is the power profile
at time $t$, and $\delta\in[0,1)$ is the discount factor assumed
to be the same for all the UEs as in \cite{etkin2007spectrum,xiao2012dynamic}.
We also assume the channel gain to be fixed over the considered time
horizon as in \cite{ephremides1990scheduling,aggarwal2011achieving,graph-dynamic}\cite{etkin2007spectrum,xiao2012dynamic}.
We will illustrate in Section VI-D and Section \ref{sub:Robustness-to-Shadow}
that the proposed framework can be adapted to the scenarios in which
the channel conditions are time-varying.


An interference management policy $\bm{\pi}^{const}$ is a policy
based on power control \cite{Decen-pow-game}, if $\bm{\pi}(t)=\textbf{p}$
for all $t$. Write the joint throughput profile of all the UEs as
$\textbf{r}(\textbf{p})=(r_{1}(\textbf{p}),...,r_{N+M}(\textbf{p}))$.
Then the set of all joint throughput profiles achievable by policies
based on power control can be written as $\mathcal{R}^{const}=\{\textbf{r}(\textbf{p}),\;\textbf{p}\in\mathcal{P}\}$.
As we have discussed before, our proposed policy is based on MISs
of the interference graph. The interference graph $G$ has $M+N$
vertices, which are the $M+N$ cells. There is an edge between two
cells if their cross interference is high. We will describe in detail
how to construct the interference graph in Section V. Given an interference
graph, we write $\textbf{I}^{G}=\{I_{1}^{G},...,I_{s(G)}^{G}\}$ as
the set of all the MISs of the interference graph. Let $\textbf{p}^{I_{j}^{G}}$
be a power profile in which the UEs in the MIS $I_{j}^{G}$ transmit
at their maximum power levels, namely $p_{k}=p_{k}^{max}\triangleq\max\mathcal{P}_{k}\;\text{if}\;k\in I_{j}^{G}$
and $p_{k}=0$ otherwise. Let $\mathcal{P}^{MIS(G)}=\{\textbf{p}^{I_{1}^{G}},...,\textbf{p}^{I_{s(G)}^{G}}\}$
be the set of all such power profiles. Then $\bm{\pi}$ is a policy
based on MIS if $\bm{\pi}(t)\in\mathcal{P}^{MIS(G)}$ for all $t$.
We denote the set of policies based on MISs by $\Pi^{MIS(G)}=\{\boldsymbol{\pi}:\mathbb{Z}_{+}\rightarrow\mathcal{P}^{MIS(G)}\}$.
The set of joint $instantaneous$ throughput profiles achievable by
policies based on MIS is then $\mathcal{R}^{MIS(G)}=\{\textbf{r}(\textbf{p}):\;\textbf{p}\in\mathcal{P}^{MIS(G)}\}$.
We will prove in Theorem 1 that the set of joint discounted throughput
profiles achievable by policies based on MIS is $\mathcal{V}^{MIS(G)}=\text{conv}\{\mathcal{R}^{MIS(G)}\}$,
where $\text{conv}\{X\}$ representing the convex hull of set $X$.

\section{Problem Formulation}

\label{sec:prblm formuln} In this section, we formalize the interference
management policy design problem, and subsequently give a motivating
example to highlight the advantages of the proposed policy over existing
policies in solving this problem.\vspace{-1em}

\subsection{Policy Design Problem}

\label{subsec:policy design} The designer of the network  (e.g.
the network operator) aims to design an optimal interference management
policy $\boldsymbol{\pi}$ that fulfills each UE $i$'s minimum throughput
requirement $R_{i}^{min}$ and optimizes a chosen network performance
criterion $W(R_{1}(\boldsymbol{\pi}),....,R_{N+M}(\boldsymbol{\pi}))$.
The network performance criterion $W$ is an increasing function in
each $R_{i}$. For instance, $W$ can be the weighted sum of all the
UEs' throughput, i.e.$\sum_{i=1}^{N}w_{i}^{FUE}R_{i}(\boldsymbol{\pi})+\sum_{j=1}^{M}w_{j}^{MUE}R_{N+j}(\boldsymbol{\pi})$
with $\sum_{i=1}^{N}w_{i}^{FUE}+\sum_{j=1}^{M}w_{j}^{MUE}=1$ and
$w_{i}^{MUE},w_{j}^{FUE}\geq0$ . We emphasize that the higher-priority
of MUEs can be reflected by setting higher weights for the MUEs (i.e.
$w_{i}^{MUE}\geq w_{j}^{FUE}\,\forall i=\{1,\ldots,N\},\,\forall j=\{1,\ldots,M\}$),
and  by setting higher minimum throughput requirements for MUEs. The
performance criterion $W$ can also be max-min fairness (i.e., the
worst UE's throughput $\min_{i}R_{i}(\boldsymbol{\pi})$) or the proportional
fairness. The policy design problem is given as follows.
\begin{eqnarray}
\mathbf{Design\,Problem} & \max_{\boldsymbol{\pi}} & W(R_{1}(\boldsymbol{\pi}),....,R_{N+M}(\boldsymbol{\pi}))\\
 & s.t. & R_{i}(\boldsymbol{\pi})\geq R_{i}^{min},\forall i\in\{1,...,N+M\}\nonumber
\end{eqnarray}

The key steps and the challenges in solving the design problem are
as follows: 1) How to determine the set of achievable throughput profiles?
Note that the set depends on the discount factor $\delta$. It is
an open problem to determine the set of achievable throughput profiles,
even for the special case of $\delta=0$ (i.e. the set of throughput
profiles achievable by policies based on power control). 2) How to
construct the optimal policy that achieves the optimal target throughput
profile? The optimal policy again depends on $\delta$. It is much
more challenging to determine the policy for delay-sensitive applications
(i.e. $\delta<1$) than for delay-insensitive applications (i.e. $\delta\rightarrow1$),
because the optimal policy is not cyclic. 3) How to construct a distributed
policy with minimum communication overhead?

\vspace{-0.6em}

\subsection{Motivating Example\label{sub:Motivating-Example}}

We consider a network of 5 femtocells. On the left plot of Fig. 1
a), we have portrayed the interference graph of this network. Each
vertex denotes a pair of FBS and its FUE. Each edge denotes strong
local interference between the connected vertices (i.e. the distance
between the FBSs is below some threshold). The interference graph
is a pentagon, where each UE interferes only with two neighbors. We
show the partitioning of the UEs by coloring the interference graph.
There are three colors, and there is one color (i.e. black) to which
only one UE belongs. On the right plot of Fig. 1 a), we show the 5
MIS's, each of which consists of two UEs. Note that the MIS are not
disjoint. For illustrative purposes, suppose that the 5 femtocells
and their UEs are symmetric, in the sense that  all the UEs have maximum
transmit power of 30mW, direct channel gain of 1, cross channel gain
of 0.25 between the neighbors, noise power at the receiver of 2mW,
minimum throughput requirement of 1.2 bits/s/Hz, and discount factor
of 0.8 representing delay sensitivity. For simplicity, we set the
cross channel gain between non-neighbors to be 0.

We compare our proposed policy against the following policies discussed
in Section \ref{sec:related works}:
\begin{itemize}
\item Policies based on power control \cite{Decen-pow-game}\cite{tan2011spectrum},
in which each UE chooses a constant (time-invariant) power level all
the time.
\item Coloring-based TDMA policies \cite{ramanathan1993scheduling}\cite{pateromichelakis2012dynamic},
in which the UEs are partitioned into mutually exclusive subsets by
coloring the interference graph; in each time slot, all the UEs of
one color are chosen to transmit. In this example, 3 colors are required
and there will always exist a color to which only one UE belongs.
Hence, the average number of active UEs in each time slot is less
than 2. Note that the optimal performance of coloring based frequency
reuse policies is the same as the optimal performance that can be
attained by any coloring based TDMA of any arbitrary cycle length.
This is due to the fact that FDM and TDM are equivalent provided the
frequency/time can be divided arbitrarily.
\item Cyclic MIS-based TDMA policies \cite{ephremides1990scheduling}\cite{aggarwal2011achieving},
in which different MISs of UEs are scheduled in a cyclic manner. In
this example, there are 5 MISs, each of which consists of 2 UEs. Hence,
the average number of active UEs in each time slot is 2. This is the
major reason why MIS-based TDMA policies are more efficient than coloring-based
TDMA policies. To completely specify the policy we must also specify
a cycle length and order of transmissions; note that the efficiency
of the policy will depend on the cycle length due to delay sensitivity.
\end{itemize}
We illustrate the performance of the above policies vs the proposed
policy in Table 1. Remarkably, the proposed policy is not only much
more efficient than existing policies, it is much easier to compute.
To compare with constant policies, note simply that finding the optimal
constant policy is intractable \cite{tan2011spectrum} in general,
because the optimization problem is non-convex due to the mutual interference.
To compare with different classes of TDMA policies, note that for
(coloring-based and MIS-based) cyclic TDMA policies, the complexity
of finding the optimal cyclic policy of a given length grows exponentially
with the cycle length (and exponentially with the number of MISs when
the cycle length is large enough for reasonable performance). To get
a hint of why this is so, note that in a cyclic policy, the UE\textquoteright s
performance is determined not only by the number of TXOPs in a cycle
but also by the positions of the TXOPs since UEs are discounting their
future utilities (due to delay sensitivity). Thus, it is not only
the length of the cycle that is important but also the ordering of
transmissions within each cycle. For instance, for the 5-UE case above,
achieving performance within 10\% of the optimal nonstationary policy
requires that the cycle length L be at least 7, and so requires searching
among the thousands (16800) \footnote{We compute the number of nonstrivial schedules by exhaustively searching
among all the possible policies.} of different nontrivial schedules (the schedules in which each UE
transmits at least once in each cycle) of cycle length 7. Even this
small problem is computationally intensive. For a moderate number
of 10 femtocells, assuming a completely connected interference graph
which has 10 MISs, and a cycle length of 20, we need to search more
than ten billion (i.e. $10^{10}$) non-trivial schedules \textendash{}
a completely intractable problem. %
{}
\begin{table}
{\scriptsize{}}\global\long\def\arraystretch{1.0}
{\scriptsize{} \protect\caption{Comparisons against spatial reuse TDMA based policies}
}{\scriptsize \par}

\centering{}{\scriptsize{}\label{table:5user_illun} \centering }%
\begin{tabular}{|l|l|l|}
\hline
{\scriptsize{}Policies } & {\scriptsize{}Max-min throughput (bits/s/Hz) } & {\scriptsize{}Performance Gain \% }\tabularnewline
\hline
{\scriptsize{}Optimal constant power} & {\scriptsize{}1.32} & {\scriptsize{}21.2\%}\tabularnewline
\hline
{\scriptsize{}Optimal Coloring TDMA (arbitrary L) } & {\scriptsize{}1.33 (Upper Bound)} & {\scriptsize{}20.3 \% }\tabularnewline
\hline
{\scriptsize{}Optimal MIS TDMA (L=5) } & {\scriptsize{}1.36 } & {\scriptsize{}17.6 \% }\tabularnewline
\hline
{\scriptsize{}Optimal MIS TDMA (L=7) } & {\scriptsize{}1.49 } & {\scriptsize{}7.8 ~\% }\tabularnewline
\hline
{\scriptsize{}Optimal Proposed } & {\scriptsize{}1.60 } & {\scriptsize{}-- }\tabularnewline
\hline
\end{tabular}{\scriptsize \par}
\end{table}

\begin{center}
\begin{figure}
\begin{centering}
\includegraphics[width=5.8in]{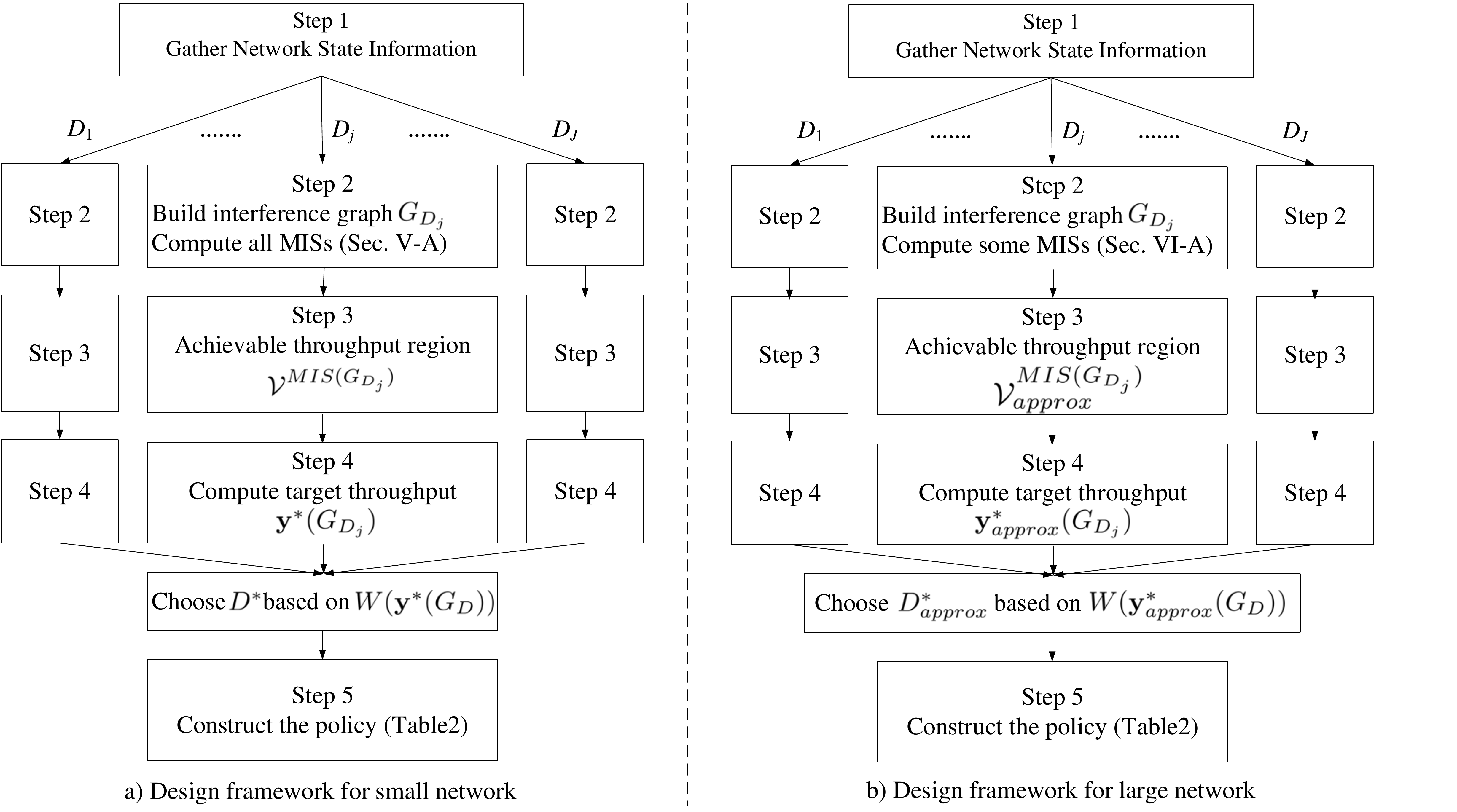}
\par\end{centering}

\protect\caption{\label{fig:design_Frmwrk} Steps in the design framework.}
\end{figure}

\par\end{center}

\section{Design Framework}

\label{sec:design framework} In this section, we develop the general
design framework for solving the design problem (1). We will provide
sufficient conditions under which our proposed framework is optimal,
and demonstrate a wide variety of networks that fulfill the sufficient
conditions.

\subsection{Description of the Proposed Design Framework\label{sub:Steps-in-the}}

The proposed methodology for solving the design problem consists of
5 steps which are illustrated in Fig. \ref{fig:design_Frmwrk} %
. We describe them in detail as follows.

\subsubsection{$\textbf{Step 1. The Designer Gathers Network Information}$}

The designer is informed by each BS $i$ of the minimum throughput
requirement $R_{i}^{min}$ of its UE, the channel gain from each UE
$j$ to its receiver $g_{ji}$, its UE's maximum transmit power level
$p_{i}^{max}$, the noise power level at its receiver $\sigma_{i}^{2}$,
and its location as in \cite{graph-dynamic}\cite{lee2010interference}.
Such information is sent to the designer via the backhaul link. In
some circumstances, the information about the location of FBSs is
available to the femtocell gateways \cite{lee2010interference}, who
can send this information to the designer.

\subsubsection{$\textbf{Step 2. The Designer Constructs the Interference Graph and Computes the MISs}$}

The designer constructs the interference graph using the information
of cell locations obtained in Step 1. Specifically, it uses a distance
based threshold rule as in \cite{huson1995broadcast}\cite{hale1980frequency}
to construct the graph: there is an edge between two cells if the
distance between BSs in these two cells is smaller than a threshold
$D$ \footnote{Note that the interference actually depends on the distance between
a BS and a UE in another cell, instead of the distance between two
BSs. When the distance from a BS to its UE is small, then the distance
between BSs is an accurate representation of interference.} . Given the threshold $D$, we denote the resulting graph by $G_{D}$,
and the set of its MISs by $\textbf{I}^{G_{D}}$, which can be calculated
as in \cite{johnson1988generating}. We assume that the distance threshold
$D$ is fixed for now, and will discuss how to select the threshold
in the next subsection.

\subsubsection{$\textbf{Step 3. The Designer Characterizes Achievable Throughput}$}

Based on the MISs ~computed in Step 2, the designer identifies the
set $\mathcal{V}^{MIS(G_{D})}(\delta)$ of throughput vectors achievable
by MIS-based policies. Note that $\mathcal{V}^{MIS(G_{D})}(\delta)$
depends on the discount factor. Recall that $\mathcal{R}^{MIS(G_{D})}=\{\textbf{r}(\textbf{p}):\;\textbf{p}\in\mathcal{P}^{MIS(G_{D})}\}$
is the set of $\emph{instantaneous}$ throughput profiles achievable
by MIS-based policies in $\Pi^{MIS(G_{D})}$. The theorem below proves
that $\mathcal{V}^{MIS(G_{D})}(\delta)$ is a convex hull of $\mathcal{R}^{MIS(G_{D})}$,
i.e. $\mathcal{V}^{MIS(G_{D})}$ when the discount factor $\delta\geq1-\frac{1}{s(G_{D})}$,
where $s(G_{D})$ is the number of MISs in the interference graph
$G_{D}$.

\textbf{Theorem 1:} Given the interference graph $G_{D}$, for any
$\delta\geq\bar{\delta}=1-\frac{1}{s(G_{D})}$,  the set of throughput
profiles achieved by MIS-based policies is $\mathcal{V}^{MIS(G_{D})}(\delta)=\mathcal{V}^{MIS(G_{D})}$.

See the detailed proof in the Appendix at the end.

\textbf{Proof Sketch: }The main step involved in proving the above
is to derive the conditions on the discount factor such that each
throughput vector in $\mathcal{V}^{MIS(G_{D})}$ can be decomposed
into a current throughput vector which belongs to $\mathcal{R}^{MIS(G_{D})}$
and a continuation throughput which belongs to $\mathcal{V}^{MIS(G_{D})}$.
To derive the conditions, we show that for any vector in $\mathcal{V}^{MIS(G_{D})}$
there exists at least one throughput vector in $\mathcal{R}^{MIS(G_{D})}$
to decompose the vector. Since the continuation throughput also belongs
to $\mathcal{V}^{MIS(G_{D})}$, it can be decomposed as well in a
similar fashion. Hence, all the vectors in $\mathcal{V}^{MIS(G_{D})}$
are achievable.

Theorem 1 is important, because it analytically characterizes the
set of throughput profiles achievable by MIS-based policies, and gives
us the requirements that need to be fulfilled by the discount factor.

\subsubsection{$\textbf{Step 4. The Designer Determines the Optimal Target Weights}$}

Among all the achievable throughput profiles identified in Step 3,
the designer selects the target throughput profile in order to optimize
the network performance. Note that each UE $i$'s average throughput
$R_{i}$ can be expressed as a convex combination of the instantaneous
throughput vectors achieved by MIS-based policies (i.e. the throughput
vectors in $\mathcal{R}^{MIS(G_{D})}$). Thus, determining the optimal
target vector and its corresponding coefficient can be formulated
as the following optimization problem :
\begin{align}
\max_{\boldsymbol{y},\boldsymbol{\alpha}}\,\, & W(y_{1}(G_{D}),...,y_{N+M}(G_{D}))\nonumber \\
s.t.\,\, & y_{i}(G_{D})\geq R_{i}^{min},\;\forall i\in\{1,...,N+M\}\nonumber \\
 & y_{i}(G_{D})=\sum_{j=1}^{s(G_{D})}\alpha_{j}r_{i}(\textbf{p}^{I_{j}^{G_{D}}}),\forall i\in\{1,....,N+M\}\\
 & \sum_{j=1}^{s(G_{D})}\alpha_{j}=1,\,\alpha_{j}\geq0,\;\forall j\in\{1,..,s(G_{D})\}\nonumber
\end{align}
 The above optimization problem is a convex optimization problem and
is easy to solve if $W$ is concave (e.g. weighted sum throughput
or max-min fairness). The resulting optimal target vector and its
corresponding coefficient is given as $\textbf{y}^{*}(G_{D})=[y_{1}^{*}(G_{D}),....,y_{N+M}^{*}(G_{D})]$
and $\boldsymbol{\alpha}^{*}(G_{D})=[\alpha_{1}^{*}(G_{D}),....,\alpha_{s(G_{D})}^{*}(G_{D})]$
respectively. Note that the optimal value depends on the interference
graph $G_{D}$ which we assume to be fixed in this section. The optimal
coefficient for the $i^{th}$ MIS $I_{i}^{G_{D}}$, i.e $\alpha_{i}^{*}(G_{D})$
can be interpreted as the fraction of time for which $I_{i}^{G_{D}}$transmits.

\subsubsection{$\textbf{Step 5. Each UE Implements the Policy Distributedly to Achieve the Target}$}

The designer informs each UE $i$ of the optimal coefficients, i.e.
$\boldsymbol{\alpha}^{*}(G_{D})$ and the indices of MISs that UE
$i$ belongs to. The designer can send the above information to each
BS $i$, who will forward the information to its UE. Each UE $i$
executes the policy in Table 2. The policy in Table 2 leads to a non-stationary
scheduling of the MISs. Note that each UE $i$ computes its own policy
online $\emph{without}$ information exchange. Hence, the computed
policy is implemented in a decentralized manner by the UEs. Next we
state the condition under which the policy indeed converges to the
target vector $\textbf{y}^{*}(G_{D})$.

\textbf{Theorem 2:} For any $\delta\geq\bar{\delta}=1-\frac{1}{s(G_{D})}$,
the policy computed in Table 2 achieves the target throughput profile
$\textbf{y}^{*}(G_{D})$.

See the detailed proof in the Appendix at the end.

\textbf{Proof Sketch: }We show that when $\delta\geq\bar{\delta}=1-\frac{1}{s(G_{D})}$,
the policy developed in Table 2 ensures that the decomposition property
given in Proof Sketch of Theorem 1 is satisfied in each time slot.
This is used to show that the distance from the target, $\textbf{y}^{*}(G_{D})$
strictly decreases in each time slot.

We briefly discuss the intuition behind our proposed policy. We determine
which MIS to transmit based on a metric that can be interpreted as
the ``fraction of time slots allocated to an MIS in the future'':
the MIS that has the maximum fraction of time slots in the future,
i.e. the highest metric, will transmit at the current time slot. The
metric is updated in each time slot as follows: the fraction of time
slots for the MIS who has just transmitted will decrease, and those
of the other MISs will increase.

\subsection{Constructing Optimal Interference Graphs\label{sub:Constructing-Optimal-Interference}}

In Step 2 of the design framework, we construct the interference graph
by comparing the distances between two BSs with a threshold $D$.
Here we show how to choose the optimal threshold $D^{*}$ and hence
the optimal interference graph $G_{D^{*}}$, based on which the proposed
policy achieves the highest network performance achievable by any
MIS based policy in $\Pi^{MIS(G_{D})}$. Formally, the designer chooses
the optimal threshold $D^{*}$ that results in the optimal interference
graph $G_{D^{*}}=\text{arg}\max_{G_{D}\in\mathcal{{G}}}W(\textbf{y}^{*}(G_{D}))$,
where $\mathcal{G}$ is the set of all possible interference graphs
constructed based on the distance rule. The designer solves the above
optimization problem by performing Steps 2-4 for each of the $|\mathcal{G}|=J$
interference graphs as shown in Fig. 2. and chooses the optimal one.
Note that the number $|\mathcal{G}|$ of all such interference graphs
is finite and upper bounded by $\frac{(M+N)\cdot(M+N-1)}{2}+1$, because
the number of different distances between BSs is finite and upper
bounded by $\frac{(M+N)\cdot(M+N-1)}{2}+1$. Note that the Steps 3-5
of our design framework can be used for any given interference graph,
which is not necessarily constructed based on the distance based threshold
rule. We assume a distance based threshold rule as a concrete example,
in order to describe how to choose the optimal interference graph.

\subsection{Optimality of the Proposed Design Framework}

Our proposed design framework first constructs the interference graph
based on the distances between BSs, and then schedules the MISs of
the constructed interference graph. Then our proposed policy let the
UEs in the scheduled MIS to transmit at their maximum power levels.
To some extent, the interference graph is a binary quantization of
the actual interference (i.e. ``no interference'' among non-neighbors
and ``strong interference'' among neighbors). Hence, the performance
of the proposed policy depends crucially on how close the interference
graph is to the actual interference pattern. If we choose a smaller
threshold $D$, the interference graph will have fewer edges, the
non-neighboring UEs will have higher cross channel gains. Hence, the
UEs in a MIS may experience high accumulative interference from the
non-neighbors. If we choose a higher threshold $D$, the interference
graph is more conservative and will have more edges. Hence, some UEs
outside a MIS may cause low interference and should be scheduled together
with the UEs in the MIS. Our proposed policy will achieve performance
close to optimality, if the interference graph is well constructed
such that: 1) neighbors have strong interference, and 2) non-neighbors
have weak interference. Next, we analytically quantify the above intuition
and provide rigorous conditions for the optimality of the proposed
design framework.

Let $W^{*}$ denote the optimal network performance, namely the optimal
value of the design problem (1) with the performance criterion being
the weighted sum throughput. We give conditions under which the proposed
policy can achieve within $\epsilon$ of the optimal performance $W^{*}$.
We first quantify strong local interference among neighbors as follows.
Define $r_{i}^{\prime}(\textbf{p})=\log_{2}(1+\frac{g_{ii}p_{i}}{\sum_{j\in\mathcal{N}_{i}(G_{D})}g_{ji}p_{j}+\sigma_{i}^{2}})$,
where $\mathcal{N}_{i}(G_{D})$ is the set of neighbors of $i$ in
$G_{D}$ and let $\mathcal{R}_{a}^{const}=\{r_{i}^{'}(\textbf{p}),\;\textbf{p}\in\mathcal{P}\}$
, $\mathcal{R}_{a}^{MIS(G_{D})}=\{r_{i}^{'}(\textbf{p}),\;\textbf{p}\in\mathcal{P}^{MIS(G_{D})}\}$
and $\mathcal{V}_{a}^{MIS(G_{D})}=\text{conv}\{\mathcal{R}_{a}^{MIS(G_{D})}\}$.
Note that $r_{i}^{\prime}(\textbf{p})$ is not the actual throughput
$r_{i}(\textbf{p})$, because we do not count the interference from
non-neighbors in $r_{i}^{\prime}(\textbf{p})$.

\textbf{Definition 1 (Strong Local Interference):} The interference
graph $G_{D}$ exhibits${\it \,\textit{Strong}\,}$$\textit{Local}\,$
$\textit{Interference}\,$(SLI) if $\mathcal{V}_{a}^{MIS(G_{D})}$
dominates $\mathcal{R}_{a}^{const}$, in the sense that every throughput
profile in $\mathcal{R}_{a}^{const}$ is weakly Pareto dominated by
a throughput profile in $\mathcal{V}_{a}^{MIS(G_{D})}$.

Definition 1 states that for an interference graph with SLI, it is
more efficient to use MIS-based policies than constant power control
policies. Next, we quantify the weak interference among non-neighbors.

\textbf{Definition 2 (Weak Non-neighboring Interference):} The interference
graph $G_{D}$ has $\epsilon-\textit{Weak}$ ${\it \textit{Non-neighboring}}$$\,\textit{Interference}$
($\epsilon$-WNI) if each UE $i$'s maximum interference from its
non-neighbors is below some threshold, namely $Int_{i}^{max}(G_{D})=\sum_{j\not\in\mathcal{N}_{i}(G_{D}),j\not=i}g_{ji}p_{j}^{max}\leq(2^{\epsilon}-1)\sigma_{i}^{2},\,\forall i\in\{1,...,N+M\}$.

Now we state Theorem 3 which uses the above two definitions to ensure
optimality.

\textbf{Theorem 3:} If the constructed interference graph $G_{D^{*}}$
exhibits SLI and $\epsilon$-WNI, then the proposed policy computed
through Steps 1-5 of Subsection V-A achieves within $\epsilon$ of
the optimal network performance $W^{*}$.

See the Appendix at the end for the detailed proof.

\textbf{Proof Sketch:} The set of throughput vector achievable by
any policy is $conv\{\mathcal{R}^{const}\}$. Denote the optimal throughput
vector by $\textbf{v}^{*}\in conv\{\mathcal{R}^{const}\}$, namely
$W(\mathbf{v}^{*})=W^{*}$. There must exist a vector $\textbf{\ensuremath{\tilde{\textbf{v}}}}\in conv\{\mathcal{R}_{a}^{const}\}$
such that $\textbf{\ensuremath{\tilde{\textbf{v}}}}\geq\textbf{v}^{*}$,
because we do not count the interference from non-neighbors when we
calculate $r_{i}^{\prime}(\textbf{p})\in\mathcal{R}_{a}^{const}$.
SLI indicates that there exists a vector $\textbf{v}^{\prime}\in\mathcal{V}_{a}^{MIS(G_{D^{*}})}$
such that $\textbf{v}^{\prime}\geq\textbf{\ensuremath{\tilde{\textbf{v}}}}\geq\textbf{v}^{*}$.
This condition implies that if hypothetically there was zero interference
from non-neighbors, then MIS based policies will achieve the optimal
throughput vector. However, since there is interference from non-neighbors,
we use $\epsilon$-WNI to bound the loss in throughput caused by the
interference from non-neighbors. ~Using $\epsilon$-WNI we can find
a throughput profile $\textbf{v}\in\mathcal{V}^{MIS(G_{D^{*}})}$
which is within $\epsilon$ from $\textbf{v}^{\prime}\in\mathcal{V}_{a}^{MIS(G_{D^{*}})}$.
Hence, we have $\textbf{v}^{'}\geq\textbf{v\,\ensuremath{\geq}\,\textbf{v}}^{'}-\epsilon$
and $\text{v}_{i}\geq R_{i}^{min}-\epsilon$. This shows that we can
achieve a throughput vector that is $\epsilon$ close to the optimal
one, i.e. $\textbf{v\,\ensuremath{\geq}\,\textbf{v}}^{*}-\epsilon$
.

\textit{Example:} Consider 3 UEs and their corresponding FBS located
on 3 different floors as shown in Fig. 3. Each UE can transmit at
a maximum power of $100$ mW. The channel model for determining the
 gain from a UE $i$ to BS $j$, which includes the attenuation from
the floor, is set based on \cite{seidel1992914}. Specifically, we
have $G_{ii}=0.5$, $G_{ji}=0.25$ for $|j-i|=1$, $G_{ji}=0.0032$
for $|j-i|=2$, and the noise power of 2 mW. We aim to maximize the
average throughput while fulfilling a minimum throughput requirement
of 1.2 bits/s/Hz for each FUE. Under three different thresholds $D$,
we have the following three  interference graphs (there are only
three interference graphs because there are only three different values
of distance between the BSs): 1) the triangle graph $\{D\geq4m\}$,
2) the chain graph $\{2m\leq D<4m\}$ and 3) the edge-free graph $\{0m\leq D<2m\}$.
For each of these graphs, we apply the design framework described
in Subsection V-A to obtain the corresponding policy, and achieve
the following average throughput: 1) 1.56 bits/s/Hz 2) 2.7 bits/s/Hz
and 3) 1.5 bits/s/Hz. Hence, the chain graph is the optimal choice
among the three graphs. Also the chain graph exhibits SLI as illustrated
in Fig. 3. and also exhibits $\epsilon-$WNI for $\epsilon=0.2$.
Hence, the proposed policy calculated based on the chain graph yields
an average throughput within $\epsilon=0.2$ of the optimal solution
$W^{*}$ to the design problem in (1) (i.e. $W^{*}\leq2.9$ bits/s/Hz).

\begin{figure}
\label{fig:3user} \centering \includegraphics[width=3.5in]{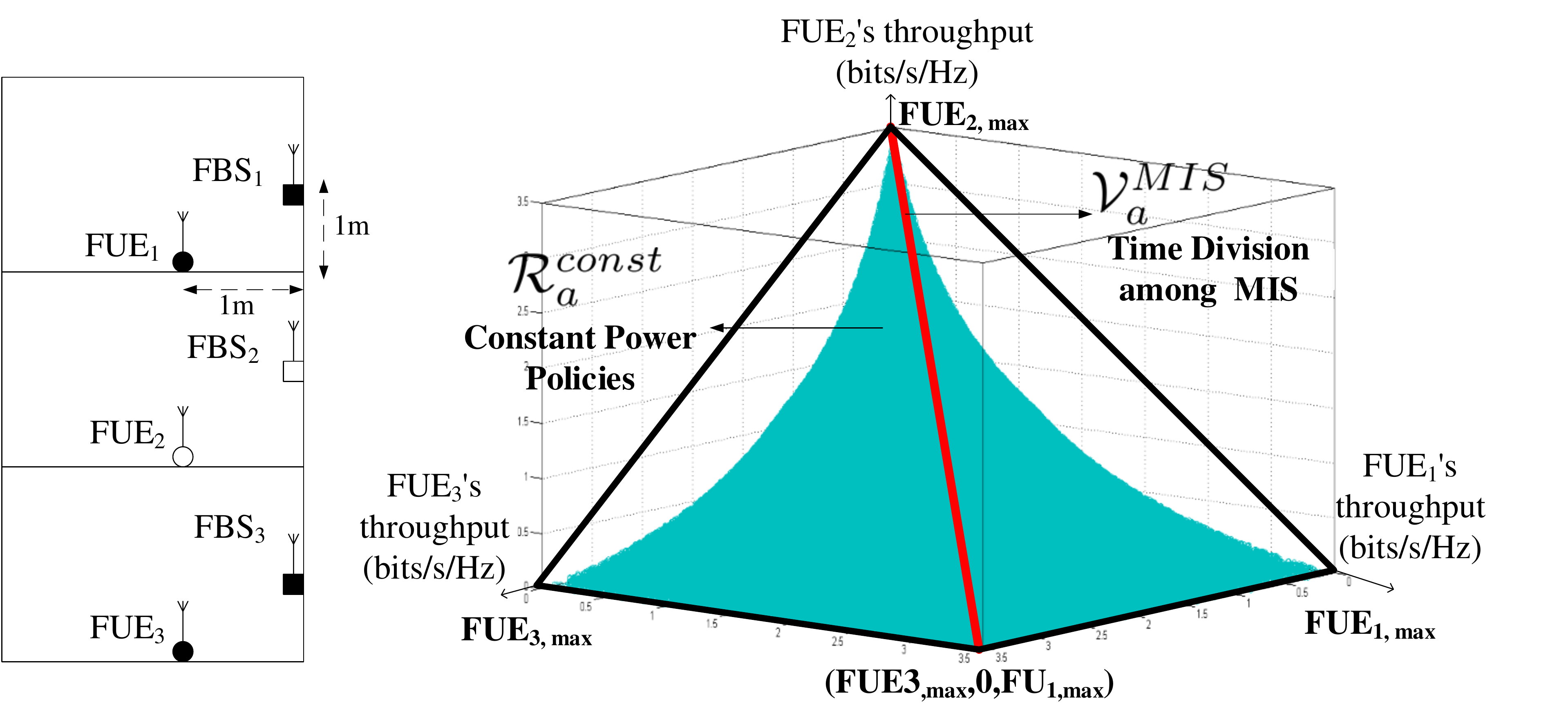}
\protect\caption{An example to illustrate the optimality of proposed framework.}

\end{figure}

\subsection{Complexity for computing the policy}

We only compare the computational complexity of the proposed policies
against cyclic MIS-based TDMA policies, since determining the optimal
constant power based policy is a non-convex problem and has been shown
to be NP-hard \cite{tan2011spectrum}. We compare the two for a given
interference graph $G_{D}$. Both the optimal cyclic MIS TDMA and
the proposed policy need to compute the set of MISs. Determining all
the MISs  in general requires large computational cost\cite{johnson1988generating}.
However, the computational complexity is acceptable if the network
is small, or if the number of MIS, $s(G_{D})=\mathcal{O}((N+M)^{c}),\;c>1$
is bounded by a polynomial function in the number of vertices in $G_{D}$.
We will develop an approximate algorithm to compute only a subset
of MISs within polynomial time and with performance guarantees in
Section \ref{sec:low complexity implem}. In our framework, the
remaining amount of computation (other than computing MISs) is dominated
by the amount of computation performed in Step 4, because in Step
5, the policy is computed online with a small amount $\mathcal{O}(s(G_{D}))$
of computations per time slot. In Step 4, we solve the optimization
problem in (5) with the objective function $W$ and linear constraints.
When $W$ is linear (e.g. weighted sum throughput) or is the minimum
throughput of any UE (in which case the problem can be transformed
into a linear programming), the worst-case computational complexity
for solving (5) is $\mathcal{O}((s(G_{D})+N+M)^{3.5}B^{2})$\cite{karmarkar1984new}
where $B$ is the number of bits to encode a variable. In contrast,
the complexity of computing the optimal cyclic MIS-based TDMA policy
of cycle length $L$ scales by $[s(G_{D})]^{L}$. The complexity quickly
becomes intractable when cycle lengths are moderately higher than
$N+M$, which is usually needed for acceptable performance. In summary,
the complexity of computing our policies is much lower than that of
computing cyclic MIS-based TDMA policies.

\subsection{Impact of the density of femtocells and macrocells}

The density of the network is defined as the average number of neighbors
of a UE in the interference graph. To obtain sharp analytical results,
we restrict our attention to a class of interference graphs with $N+M$
vertices and $H$ cliques of the same size. Note that a clique is
a subset of vertices, where any two vertices are connected. Assuming
that no two cliques are connected, we can compute the density as $d=\frac{{N+M}}{H}-1$.
When the total number $N+M$ of UEs remains the same and the density
$d$ increases, the number $H$ of cliques will decrease. Since the
vertices in a MIS can only come from different cliques, the number
of MISs decreases as $H$ decreases. As a result, the complexity of
the policy will decrease. When the density increases, the multi-user
interference increases, leading to a decrease in the throughput and
in the network performance.

\section{Efficient Interference Management With Provable Performance Guarantees
for Large-Scale Networks}

\label{sec:low complexity implem}

\subsection{Efficient Computation of A Subset of MISs\label{sub:Efficient-Computation-of}}

In our design framework proposed in Section \ref{sec:design framework},
we require the designer to compute all the MISs in Step 2. However,
computing all the MISs is in general  computationally
prohibitive for large networks. We propose an approximate algorithm
to compute a subset of MISs for a given interference graph $G_{D}$
in polynomial time and provide performance guarantees for our algorithm.
Note that the graph $G_{D}$ belongs to the class of unit-disk graphs
\cite{marathe1995simple}.

The subset of MISs are computed as follows.

i). $\textit{Approximate Vertex Coloring:}$ The designer first colors
the vertices \footnote{In minimum vertex coloring the objective is to use minimum number
of colors and each vertex has to be assigned atleast one color and
no two neighbors are assigned the same color. } of interference graph $G_{D}$ using the approximate minimum vertex
coloring scheme in \cite{marathe1995simple}. Let $\mathcal{C}_{1}=\{1,...,C(G_{D})\}$
be the indices of the colors. It is proven in \cite{marathe1995simple}
that the number of colors used is bounded by $C^{*}(G_{D})\leq C(G_{D})\leq3C^{*}(G_{D})\;\text{where}\;C^{*}(G_{D})\;$is
the minimum number of colors that can be used to color the vertices
of $G_{D}$. %

ii). $\textit{Generating MISs in a Greedy Manner:}$ Each color $i$
corresponds to an independent set $I_{i}^{'}$. For each independent
set $I_{i}^{'}$, the designer adds vertices in a greedy fashion until
the set is maximally independent. The procedure is described in Table
3. Let the output MIS obtained from Table 3 be $I_{k(i)}^{G_{D}}$,
where $k(i)$ is the index of the MIS in the original set of MISs
~$\textbf{I}^{G_{D}}$. We write all the MISs augmented from $I_{i}^{'},\,i=1,\ldots,C(G_{D})$
as $\{I_{k(1)}^{G_{D}},...I_{k(C(G_{D}))}^{G_{D}}\}.$

iii). $\textit{Generating the Approximate Maximum Weighted MIS}:$
Define a weight corresponding to each UE/vertex $i$ as $\bar{w_{i}}=r_{i}^{max}$,
where $r_{i}^{max}$ is the maximum throughput achievable by UE $i$
when all the other UEs do not transmit. Given these weights, the designer
ideally will like to find the maximum weighted MIS, ~namely the MIS
with the maximum sum weight of its vertices. However, finding the
maximum weighted MIS is NP-hard \cite{robson1986algorithms}. Hence,
the designer will find the $\eta$-approximate maximum weighted MIS,
denoted $I_{k(C(G_{D})+1)}^{G_{D}}$, using the algorithm in \cite{nieberg2005robust}.

The set of MISs computed from the above steps is then $\textbf{I}_{approx}^{G_{D}}=\{I_{k(1)}^{G_{D}},...I_{k(C(G_{D})+1)}^{G_{D}}\}$.
Note that $\{I_{k(1)}^{G_{D}},...I_{k(C(G_{D}))}^{G_{D}}\}$ ensure
that all the UEs are included in the scheduled MISs, and $I_{k(C(G_{D})+1)}^{G_{D}}$
is included for performance improvement. Given this subset of MISs,
we can define $\mathcal{P}_{approx}^{MIS(G_{D})}=[\textbf{p}^{I_{k(1)}^{G_{D}}},...,\textbf{p}^{I_{k(C(G_{D})+1)}^{G_{D}}}]$,
$\mathcal{R}_{approx}^{MIS(G_{D})}=\{\textbf{r}(\textbf{p}),\;\textbf{p}\in\mathcal{P}_{approx}^{MIS(G_{D})}\}$
and $\mathcal{V}_{approx}^{MIS(G_{D})}=\text{conv}\{\mathcal{R}_{approx}^{MIS(G_{D})}\}$.
Let $\Pi_{approx}(G_{D})=\{\boldsymbol{\pi}:\mathbb{Z}_{+}\rightarrow\mathcal{P}_{approx}^{MIS(G_{D})}\}$
be the set of policies in which only the subset of MISs are scheduled.
Steps 3,4 and 5 of the design framework in Section \ref{sec:design framework}
are performed given this subset (See Fig. 2). The results of Theorem
1 and 2 still apply to the policies in $\Pi_{approx}(G_{D})$ and
the set of achievable throughput profiles is $\mathcal{V}_{approx}^{MIS(G_{D})}$
given the $\delta\geq1-\frac{1}{C(G_{D})+1}$ (See Corollary 1 and
2 in Appendix at the end). The target vector in $\mathcal{V}_{approx}^{MIS(G_{D})}$
and the corresponding coefficient is computed as in Step 4 of Section
\ref{sec:design framework} and is denoted as $\textbf{y}_{approx}^{*}(G_{D})$,
$\boldsymbol{\alpha}_{approx}^{*}(G_{D})$ respectively. The coefficient
vector $\boldsymbol{\alpha}_{approx}^{*}(G_{D})$ along with the indices
of the MISs that UE $i$ belongs to is transmitted to the BS $i$
as in the Step 5 of Section \ref{sec:design framework}.

The main intuition for the procedure developed above is as follows.
Steps i) and ii) find MISs that contain all the UEs, and hence ensure
that the minimum throughput requirements are satisfied. Step iii)
finds the MIS that contains UEs with higher weights to optimize performance.
Given the MISs obtained in Steps i)-iii)  the Steps 3-5 of the design
framework are performed.

\begin{table}\parbox{0.45\linewidth}{
\begin{tabular}{|l|} \hline \textbf{\scriptsize Require:}{\scriptsize{} Target weights $\boldsymbol{\alpha}^{*}(G)=[\alpha_{1}^{*}(G_{D}),....,\alpha_{s(G_{D})}^{*}(G_{D})]$}
\tabularnewline \hline \textbf{\scriptsize Initialization:}{\scriptsize{} Sets $t=0$, $\alpha_{j}=\alpha_{j}^{*}$ for all $j\in\{1,...,s(G_{D})\}$. }
\tabularnewline \hline \textbf{\scriptsize repeat}{\scriptsize{} }\tabularnewline {\scriptsize ~~~~Finds the MIS with the maximum weight}
\tabularnewline{     \scriptsize $r^{*}=\arg\max_{j\in\{1,..,s(G_{D})\}}\alpha_{j}$}
\tabularnewline {\scriptsize ~~~~}\textbf{\scriptsize if}{\scriptsize{} $i\in I_{r^{*}}^{G_{D}}$ }\textbf{\scriptsize then}{\scriptsize{} }
\tabularnewline {\scriptsize ~~~~~~~~Transmits at power level $p_{i}^{t}=p_{i}^{max}$ }\tabularnewline {\scriptsize ~~~~}\textbf{\scriptsize end~if}{\scriptsize{} }
\tabularnewline {\scriptsize ~~~~Updates $\alpha_{j}$ for all $j\in\{1,...,s(G_{D})\}$ as follows }
\tabularnewline {\scriptsize ~~~~~~~~$\alpha_{r^{*}}=\frac{\alpha_{r^{*}}-(1-\delta)}{\delta}$, }
\tabularnewline {\scriptsize ~~~~~~~~$\alpha_{j}=\frac{\alpha_{j}}{\delta}$ $\forall j\not=r^{*}$}
\tabularnewline {\scriptsize ~~~~$t\leftarrow t+1$ }
\tabularnewline \textbf{\scriptsize until}{\scriptsize{} $\varnothing$ }\tabularnewline \hline \end{tabular}\vspace{0.2em}\caption{Algorithm run by each UE }}
\hfill
\parbox{0.45\linewidth}{\begin{tabular}{|l|}\hline \textbf{\scriptsize Require:}{\scriptsize{} $V=\{1,..,N+M\}$ set of vertices, $\textbf{\ensuremath{\bar{w}}}$ }
\tabularnewline{ \scriptsize vector of weights of vertices, Independent set~$I_{i}^{'}$ and}
\tabularnewline {\scriptsize $\text{Adj}(I_{i}^{'})$ where Adj(X) is the set of neighbors of $X$}
\tabularnewline \hline \textbf{\scriptsize Initialization:}{\scriptsize{} $I_{k(i)}^{G_{D}}=I_{i}^{'}$, $\mathcal{N}_{i}^{'}=V\cap(I_{i}^{'}\cup\text{Adj}(I_{i}^{'}))^{c}$, }
\tabularnewline{\scriptsize here $(X)^{c}$ is the complement of $X$}\tabularnewline \hline \hline {\scriptsize While( $\mathcal{N}_{i}^{'}\neq\phi$) }\tabularnewline {\scriptsize $\mathcal{N}_{i}^{'}=\text{sort}(\mathcal{N}_{i}^{'})$ , sort the vertices }\tabularnewline{\scriptsize in $\mathcal{N}_{i}^{'}$ in the decreasing order of the weights $\bar{w}_{j}$ }\tabularnewline {\scriptsize $v^{'}=\mathcal{N}_{i,1}^{'}$ , here $\mathcal{N}_{i,1}^{'}$ is the first vertex in $\mathcal{N}_{i}^{'}$ }\tabularnewline {\scriptsize $I_{k(i)}^{G_{D}}=I_{k(i)}^{G_{D}}\cup v^{'}$ }\tabularnewline {\scriptsize $\mathcal{N}_{i}^{'}=\mathcal{N}_{i}^{'}\cap(\{v^{'}\}\cup\text{Adj}(\{v^{'}\})^{c}$ }\tabularnewline {\scriptsize end }  \tabularnewline \tabularnewline \hline \end{tabular}\vspace{0.2em}\caption{Algorithm run by the designer}}\end{table}

\subsection{Performance Guarantees for Large Networks\label{sub:Performance-Guarantee-for}}

In this subsection, we consider the network performance criterion
as the weighted sum throughput, and give performance guarantees for
the policy when we compute the subset of MISs by Steps i)-iii) in
the Subsection \ref{sub:Efficient-Computation-of}. Note that as we
will show in the Section \ref{sec:simulations}, the subset of MISs
perform well in large networks for other network performance metrics
as well. In particular the performance guarantee implies that the
performance scales with the optimum $W(\textbf{y}^{*}(G_{D}))$ as
the network size $N+M$ increases. Define $D_{ij}^{UE}$ as the distance
from UE-$i$ to BS-$j$. We make the following homogeneity assumption,
$p_{i}^{max}=p^{max}$, $\sigma_{i}^{2}=\sigma^{2}$, $R_{i}^{min}=R^{min}$,
$\max_{i}D_{ii}^{UE}\leq\Delta$ and $w_{i}=\frac{1}{N+M}$ \footnote{We can extend our result to a heterogeneous network with $p_{i}^{max}\geq p^{max}$,
$\sigma_{i}^{2}\leq\sigma^{2}$, $R_{i}^{min}\leq R^{min}$, $\max_{i}D_{ii}^{UE}\leq\Delta$
and $w_{i}\geq\frac{c}{N+M}$ with c as a constant. But we do not
show this general result to avoid overly complicated notations. }. Here $\Delta$ is fixed and does not depend on the size of the network.
We fix these parameters in order to understand the performance guarantee
as a function of the network size. Let the channel gain $g_{ij}=\frac{1}{(D_{ij}^{UE})^{np}}$,
where $np$ is the path loss coefficient.

We choose the trade-off variables $\rho,\zeta,\kappa$ that satisfy
$\rho+1<\min\{\frac{\log_{2}(1+\frac{p^{max}}{\Delta^{np}2^{\zeta}\sigma^{2}})}{3R^{min}},\frac{\kappa}{\zeta(1+\eta)}\log_{2}(1+\frac{p^{max}}{\Delta^{np}\sigma^{2}})\}$
and $0<\kappa<1$. Any eligible triplet $\rho,\zeta,\kappa$ will
define a class of  interference graphs that exhibit $\zeta$-WNI and
have maximum degrees upper bounded by $\rho$. Note that such interference
graphs can have arbitrarily large sizes (see the example at the end
of this subsection). Then the following theorem provides performance
guarantees for the policy described in Subsection VI-A for this class
of interference graphs.


\textbf{Theorem 4:} For any interference graph that has a maximum
degree no larger than $\rho$ and exhibits $\zeta$-WNI with $\rho+1<\min\{\frac{\log_{2}(1+\frac{p^{max}}{\Delta^{np}2^{\zeta}\sigma^{2}})}{3R^{min}},\frac{\kappa}{\zeta(1+\eta)}\log_{2}(1+\frac{p^{max}}{\Delta^{np}\sigma^{2}})\}$
the policy in Subsection \ref{sub:Efficient-Computation-of} achieves
a performance $\textbf{W(y}_{approx}^{*}(G_{D}))$ with a guarantee
that $\textbf{W(y}_{approx}^{*}(G_{D}))\geq\frac{(1-\gamma)(1-\kappa)}{(1+\eta)}\cdot\textbf{W(y}^{*}(G_{D}))$,
where $\gamma=(3(\rho+1))\frac{R^{min}}{\log_{2}(1+\frac{p^{max}}{\Delta^{np}2^{\zeta}\sigma^{2}})}$.

See the Appendix at the end for the detailed proof.

\textbf{Proof Sketch:} The condition that the graph does not have
a degree more than $\rho<\frac{\log_{2}(1+\frac{p^{max}}{\Delta^{np}2^{\zeta}\sigma^{2}})}{3R^{min}}-1$and
the $\zeta$-WNI condition ensure that the algorithm proposed in Subsection
\ref{sub:Efficient-Computation-of} yields a feasible solution satisfying
each UE's minimum throughput constraint. Also, it is shown that the
minimum coefficient/fraction of time allocated to $I_{k(C(G_{D})+1)}^{G_{D}}$is
$\alpha_{approx,k(C(G_{D})+1)}\geq(1-\gamma$). Then it is shown that
if UEs in $I_{k(C(G_{D})+1)}^{G_{D}}$ were to transmit all the time
then the competitive ratio achieved is no smaller than $\frac{1-\kappa}{(1+\eta)}$
. This combined with minimum coefficient of $I_{k(C(G_{D})+1)}^{G_{D}}$
leads to the competitive ratio guarantee of no less than $\frac{(1-\gamma)(1-\kappa)}{(1+\eta)}$.

The trade-off variables $\rho,\zeta,\kappa$ as their name suggests
provide trade-offs between how large is the class of interference
graphs for which we can provide performance guarantees, and how good
are the competitive ratio guarantees. On one hand, a higher $\kappa$
allows higher $\rho$, and higher $\rho$ and $\zeta$ allow a larger
class of graphs. On the other hand, as we can see from Theorem 4,
higher $\rho$ and $\zeta$, provided that they are eligible (higher
$\zeta$ decrease the maximum eligible $\rho$), result in higher
$\gamma$, and higher $\gamma$ and $\kappa$ give lower competitive
ratio guarantees. Hence, we can tune the design parameters to provide
different levels of competitive ratio guarantees for different classes
of interference graphs. Next, we give an example to illustrate Theorem
4.

Example: Consider a layout of FBSs in a $K\times K$ square grid,
i.e. $K^{2}$ FBSs with a distance of $5$m between the nearest FBSs,
and assume that each FUE is located vertically below its FBS at a
distance of $1$ m. Fix the parameters $p^{max}=100$ mW, $\sigma^{2}=3$
mW, $R^{min}=0.1$bits/s/Hz, $\eta=0.1$, $np=4$ and the threshold
$D=7$ m, which gives us the upper bound $\rho=4$ on the maximum
degrees. We can also verify that the interference graphs under any
number $K^{2}$ of FBSs exhibit $\zeta$-WNI with $\zeta=0.15$. Given
$\rho=4$ and $\zeta=0.15$, we choose the minimum $\kappa=0.17$,
which provides the highest competitive ratio guarantee of $0.53$.
This performance guarantee holds for any interference graph of any
size $K$.

To construct an efficient interference graph similar to Subsection
\ref{sub:Constructing-Optimal-Interference} when the number of MISs
are large, the designer must use the low complexity framework in Subsection
\ref{sub:Efficient-Computation-of}. In this case the designer computes
the subset of MISs as described in Subsection \ref{sub:Efficient-Computation-of}
and compares the optimal solution obtained to decide the best threshold
. Formally stated, the designer computes $G_{D_{approx}^{*}}=\text{arg}\max_{G\in\mathcal{G}}W(\textbf{y}_{approx}^{*}(G))$.
See Fig. 2. for a comparison of the design framework in Subsection
\ref{sub:Efficient-Computation-of} for large networks with that in
Subsection \ref{sub:Steps-in-the} for small networks.

\subsection{Complexity for computing the Subset of MISs}

We show that the proposed approximation method for computing the subset
of MISs described in Subsection \ref{sub:Efficient-Computation-of}
 has a complexity bounded by a polynomial in the number of vertices,
i.e., $\mathcal{O}((N+M)^{c}),\;c>1$. This is because Steps i). and
iii) use the algorithms developed in \cite{marathe1995simple} and
\cite{nieberg2005robust} for which the complexity has been proven
to be polynomial and ii). uses a greedy strategy in which there can
be a maximum of $N+M$ iterations since atleast one vertex is always
removed from $\mathcal{N}_{i}^{'}$ in each iteration. The worst possible
number of computations in an iteration is bounded by $(N+M)^{2}$.
Hence the upper bound of the complexity of Step ii). $\mathcal{O}((N+M)^{3}$).
Hence, the subsets of the MISs can be computed within polynomial time,
and the policy computed using this subset can guarantee a constant
competitive ratio as shown in Subsection \ref{sub:Performance-Guarantee-for}.

.

\subsection{Extensions}

\subsubsection{Construction of interference graphs based on other rules}

Our design frameworks in Section V-A and Section VI-A do not rely
on a specific method for constructing the interference graph. In Step
2 of the design frameworks (i.e., the step in which the interference
graph is constructed), we can replace our distance-based construction
of the interference graph with construction based on other criteria,
such as SINR, interference levels \cite{graph-dynamic}, etc. Then
we can use the resulting interference graph as the input to Step 3.
For construction rules based on other criteria, we can also use the
procedure described in Section V-B to optimize the construction rule
(e.g., to choose the optimal threshold of SINR or the interference
level, above which an edge is drawn between two nodes).

Note that in the design framework in Section VI, we find a subset
of MISs, instead of all MISs, because the network is large. To find
this subset, we use the coloring algorithm in \cite{nieberg2005robust},
which is known to have polynomial-time complexity for unit-disk graphs.
This is where we used the fact that the interference graph is constructed
based on distances (such that the resulting graph is a unit-disk graph).
However, we can use other polynomial-time coloring algorithms if the
interference graph is generated based on other criteria. We can use
a standard greedy coloring algorithm as in \cite{erciyes2013distributed}.
In the next step we extend the ISs obtained by coloring to MISs. We
can do this based on Step ii) in Section VI-A. The target weights
and the corresponding schedule for these MISs can be generated based
on Section VI-A. Results about the performance guarantees in terms
of competitive ratio (See Theorem 4) can also be extended to this
case.

\subsubsection{Incorporating uncertainty in channel gains}

Our design frameworks in Section V and Section VI can be extended
to the deployment scenarios in which the channel gains are not static.
For fast fading, we can replace the instantaneous throughput with
the expected instantaneous throughput in our design frameworks. For
slow fading, we can track the fading by regularly re-computing the
policy. Re-computing the entire policy every time may be costly. In
Section VII-C we show that the designer does not need to re-compute
the entire policy to get considerable gains compared to the state-of-the-art.
Specifically, the designer fixes the interference graph that is selected
in the beginning, and only re-computes the target weights rather than
re-compute the optimal interference graph and the corresponding target
weights. We also show that the performance loss incurred with respect
to the latter approach, which is based on an entire re-computation
is limited (8\%).

\subsubsection{Incorporating beamforming}

We focus on the case where each UE has one antenna. When UEs have
multiple antennas, we can easily incorporate beamforming in our framework.
Beamforming mitigates the interference among the UEs served by the
same BS. Hence, we can remove the edges between UEs in the same cell
from the interference graph. Then we can use the new interference
graph as the input to Step 3 of our design framework.

\section{Illustrative Results}

\label{sec:simulations}

\begin{figure}
\begin{minipage}[t]{0.55\textwidth}%
\includegraphics[width=3.3in,height=0.8in]{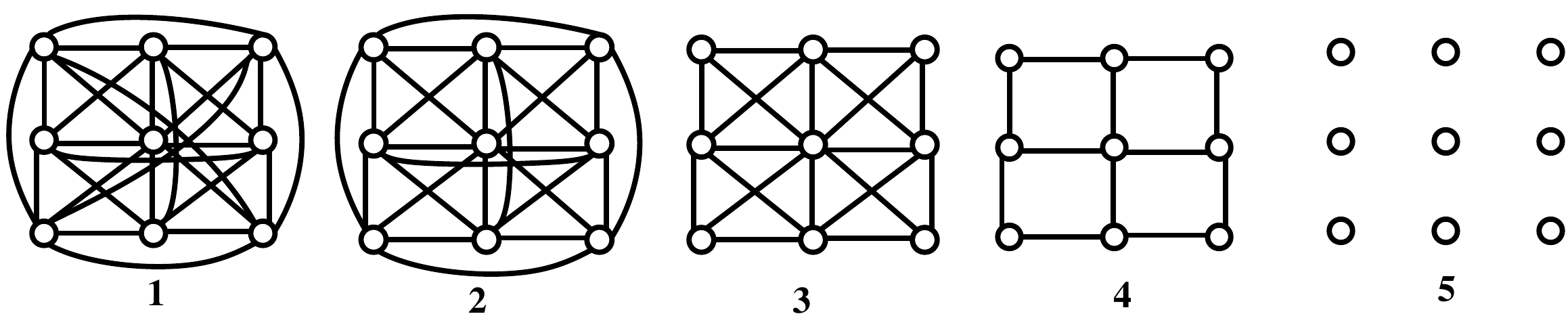}

\tiny\text{Fig. 4 a)}%
\end{minipage}%
\begin{minipage}[t]{0.35\textwidth}%
\includegraphics[width=1.7in]{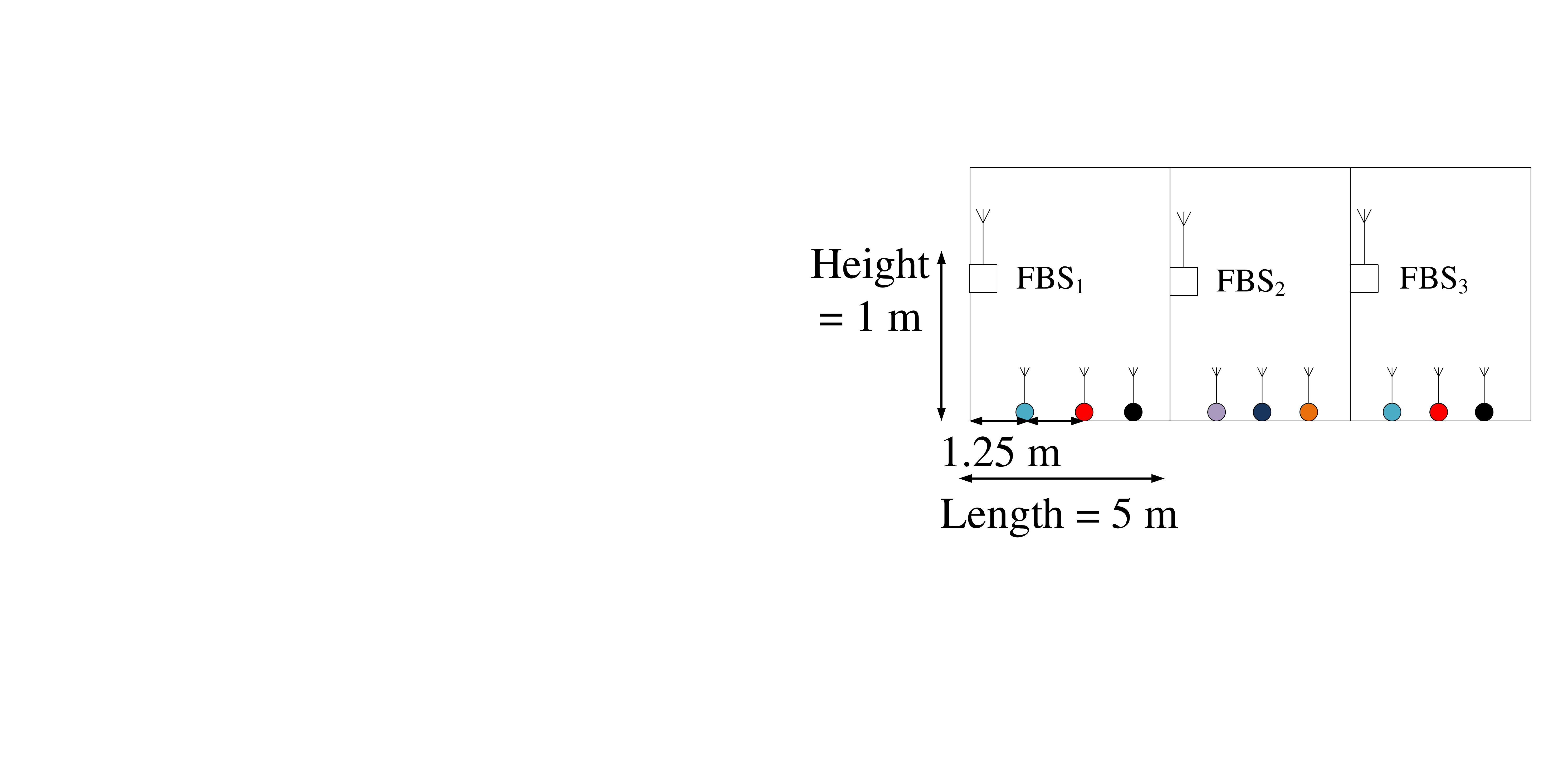}

\tiny\text{Fig. 4 b)}%
\end{minipage}\label{fig:int graph 9 user-1} \centering \protect\caption{a) Different interference graphs for the 3 x 3 BS grid, b) Illustration
of setup with 3 rooms. }

\end{figure}

In this section, we show via simulations that our proposed policy
significantly outperforms existing interference management policies
under different performance criteria. These performance gains are
obtained under varying interference levels for both small and large
networks. We also evaluate the proposed policy when the channel conditions
are time-varying due to fading. In this case, the designer ideally
needs to re-compute the optimal interference graph each time the channels
change at the cost of a higher complexity. We show the robustness
of the proposed policy when we choose a fixed interference graph regardless
of the time-varying fading.

In each setting, we compare with the state-of-the-art policies described
in Section \ref{sec:related works}, namely the constant power control
based policies and the cyclic MIS TDMA based policies. We do not compare
with coloring based TDMA/Frequency reuse policies as it was already
shown in Subsection \ref{sub:Motivating-Example} that the MIS based
TDMA policies will always lead to better network performance. Throughout
this section, we will set the discount factor as the minimum one required
when we use our original design framework in Section V (namely $\delta=1-\frac{1}{s(G_{D})}$
according to Theorems 1-2), and the minimum one required when we use
the approximate design framework for large networks in Section VI
(namely $\delta=1-\frac{1}{C(G_{D})+1}$). In this way, we evaluate
the performance of our proposed policies under the most delay-sensitive
applications.

\begin{figure}
\begin{minipage}[t]{0.45\textwidth}%
\includegraphics[width=3in]{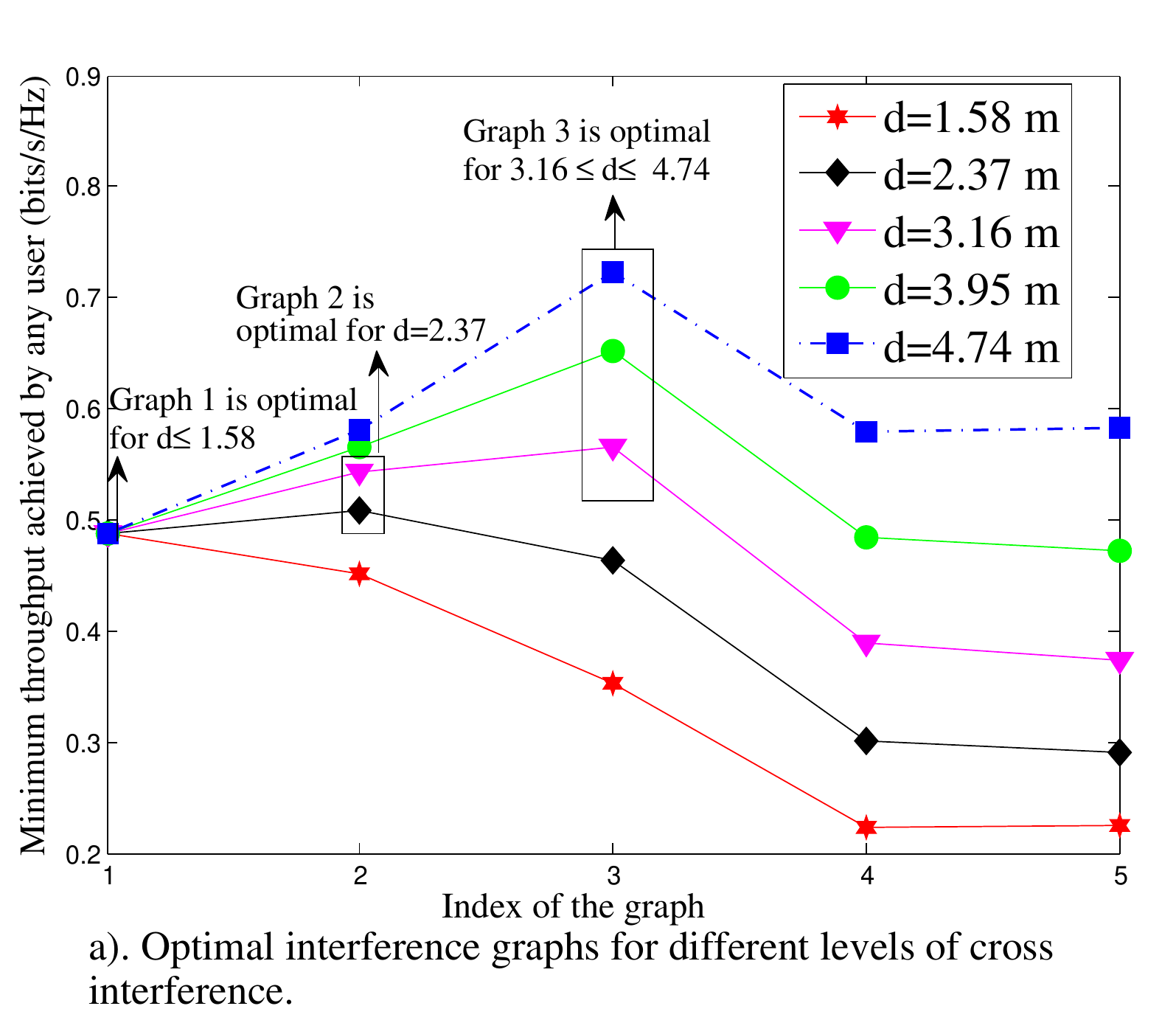}%
\end{minipage}\hspace{0.75in}%
\begin{minipage}[t]{0.45\textwidth}%
\includegraphics[width=3in]{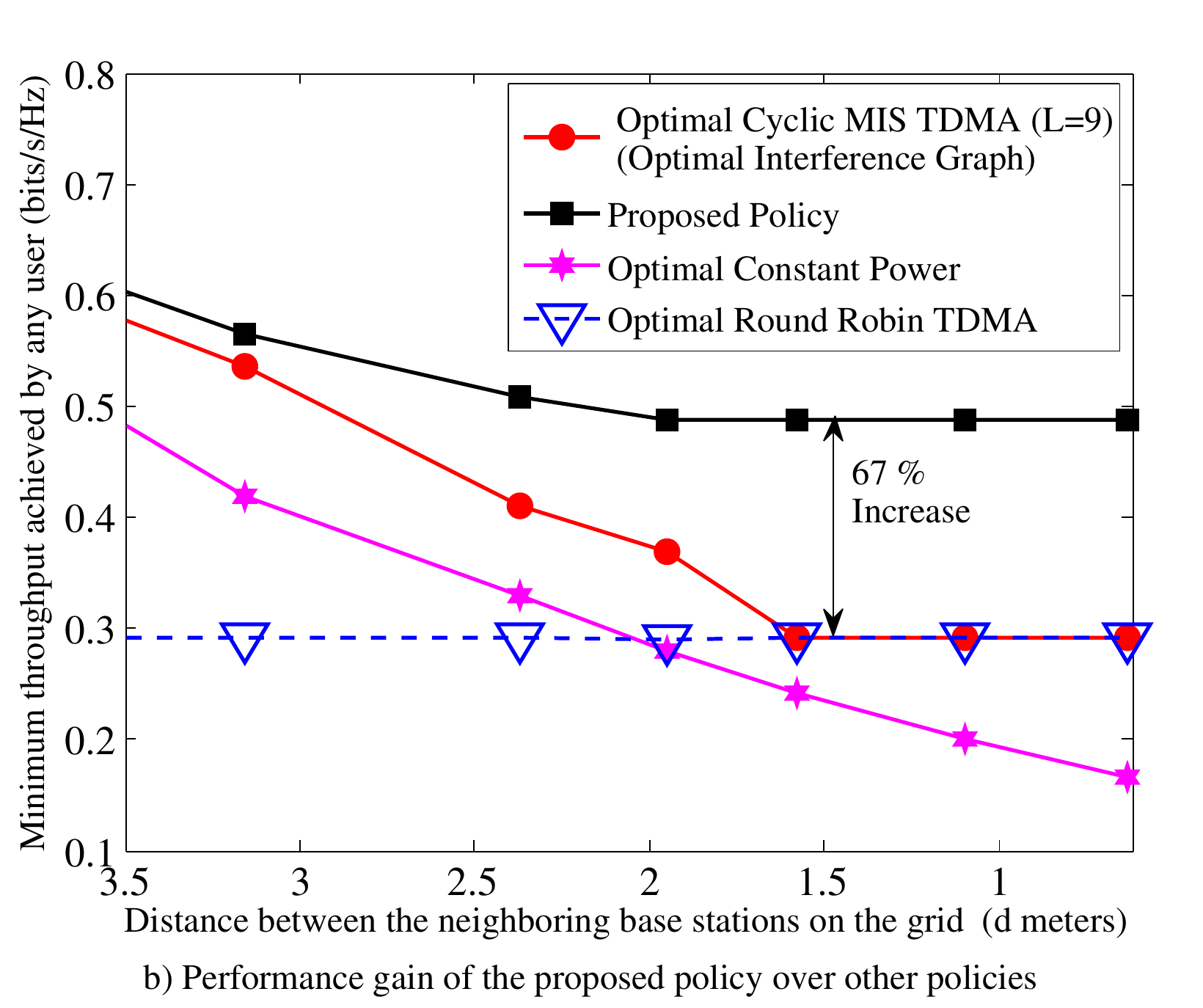}%
\end{minipage}\protect\caption{a) Optimal interference graph selection for 3 x 3 grid. b) Performance
comparison of the proposed policy for different size of the grid.}
\end{figure}

\subsection{Performance Gains Under Varying Interference Levels \label{sub:-Performance-gain}}

Consider a $3\text{x}3$ square grid of 9 BSs and corresponding UEs
with the minimum distance between any two BSs given as $d$ (see Fig.
4 a.). Each UE $i$ has $\delta=0.89$ and a maximum power of $200$
mW. Assume that the UEs an the BSs are in two horizontal hyperplanes,
and each BS is vertically above its UE with a distance of $3.16m$.
Then the distance from UE $i$ to another BS $j$ is $D_{ij}^{UE}=\sqrt{3.16^{2}+(D_{ij}^{BS})^{2}}$
, where $D_{ij}^{BS}$ is the distance between BSs $i$ and $j$.
The channel gain from UE $i$ to BS $j$ is $g_{ij}=\frac{1}{(D_{ij}^{UE})^{2}}$.
The performance criterion is the max-min fairness. Under different
thresholds $D$ chosen by the designer, there are 5 possible topologies
of the interference graph, as shown in the Fig. 4 a. For each grid
size $d$, the optimal solution to (2) is computed for each interference
graph as described in Subsection \ref{sub:Constructing-Optimal-Interference}.
Fig. 5 a). shows that under different grid sizes (i.e. different interference
levels), the optimal interference graph (i.e. the optimal threshold)
changes. As the interference level increases, the corresponding optimal
interference graph has more edges. Fig. 5 b). compare the performances
of different policies under different grid sizes $d$ (i.e. different
interference levels), where the interference graph is the optimal
one. We can see that the proposed policy achieves up to 67$\%$ performance
gain over the second best policy. Through the above results, we see
that 1) it is important to construct different interference graphs
based on the interference level, and 2) the proposed non-stationary
schedule of MISs outperform the cyclic schedules.

\subsection{Performance Scaling in Large Networks\label{sub:Performance-Scaling-in}}

We study a dense deployment scenario to evaluate the performance gain
of our proposed scheme over the state-of-the-art. We allow more than
one UE to transmit to a single BS, and will increase the number of
UEs associated with a BS. Consider the uplink of a femtocell network
in a building with 12 rooms adjacent to each other. Fig. 4 b). illustrates
3 of the 12 rooms with 3 UEs in each room. For simplicity, we consider
a 2-dimensional geometry, in which the rooms and the FUEs are located
on a line. Each room has a length of 6 meters. In each room, there
are $P$ uniformly spaced FUEs, and one FBS installed on the left
wall of the room at a height of 2m. The distance from the left wall
to the first FUE, as well as the distance between two adjacent FUEs
in a room, is $\frac{6}{(1+P)}$ meters. Based on the path loss model
in \cite{seidel1992914}, the channel gain from each FBS $i$ to a
FUE $j$ is $\frac{1}{(D_{ij}^{UE})^{2}\Delta^{n_{ij}}}$, where $\Delta=10^{0.25}$
is the coefficient representing the loss from the wall, and $n_{ij}$
is the number of walls between FUE $i$ and FBS $j$. Each UE has
a maximum transmit power level of $1000$ mW and a minimum throughput
requirement of $R_{i}^{min}=0.05$ bits/s/Hz. For each $P$, the designer
chooses the optimal threshold to construct the optimal interference
graph. Note that the UEs in the same room accessing the same BSs are
all connected to each other in the interference graph, since the distances
between their receiving BSs is $0$.

We vary the number $P$ of FUEs in each room from $5$ to $15$. We
fix the $\delta=0.97$ , i.e. the least value it can take based on
the largest number of UEs per room, i.e. $P=15$. For each $P$, the
designer constructs the optimal graph G as described in Subsection
\ref{sub:Efficient-Computation-of} using the low complexity method
as the number of UEs is large. Under all considered values of $P$,
the optimal interference graph connects all the UEs in adjacent rooms
with edges and does not connect the UEs in non-adjacent rooms. We
use the same optimal graph to compute the optimal cyclic MIS TDMA
of cycle length L. The cycle length is varied from $12$ to $58$
depending upon the number of UEs (we try to choose as large cycle
lengths as possible to maximize performance within a feasible computational
complexity). The number of non-trivial cyclic policies under different
$P$ may vary from $10^{8}$ to even more than $10^{50}$  which
renders exhaustive search to be intractable. Hence, for each $P$
we do a randomized search in $4$ million policies to search for the
optimal one. Fig. 6. compares the performance of different policies
in terms of both the max-min fairness and the sum throughput. The
constant power policy cannot satisfy the feasibility conditions for
any number of UEs in each room. The performance gain over cyclic MIS
TDMA policies increases as the network becomes larger. When there
are 15 UEs in each room, we can improve the worst UE's throughput
by $131\%$ compared to cyclic MIS TDMA policies.


\begin{figure}
\begin{minipage}[t]{0.45\textwidth}%
\includegraphics[width=3in]{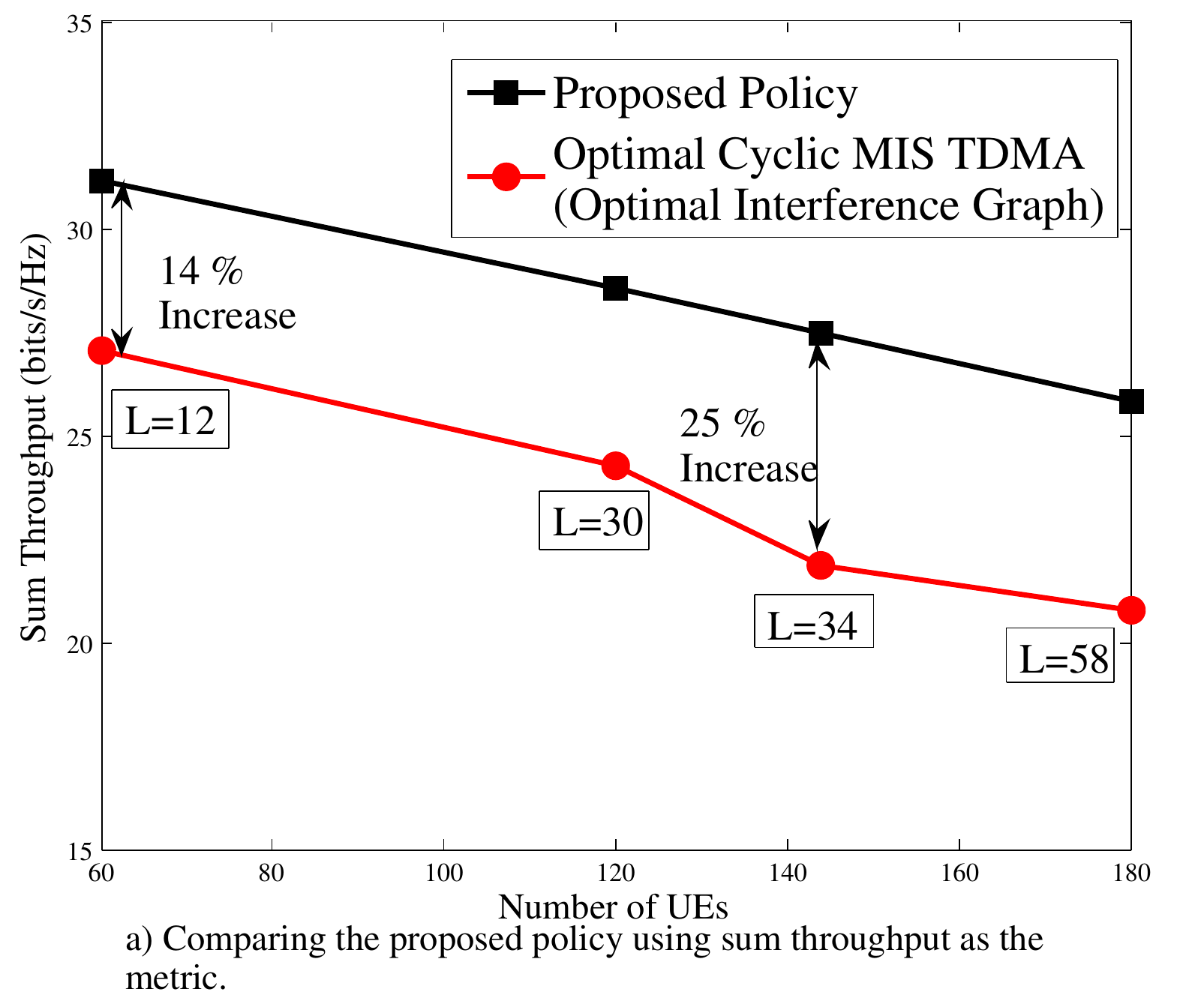}%
\end{minipage}\hspace{0.75in}%
\begin{minipage}[t]{0.45\textwidth}%
\includegraphics[width=3in]{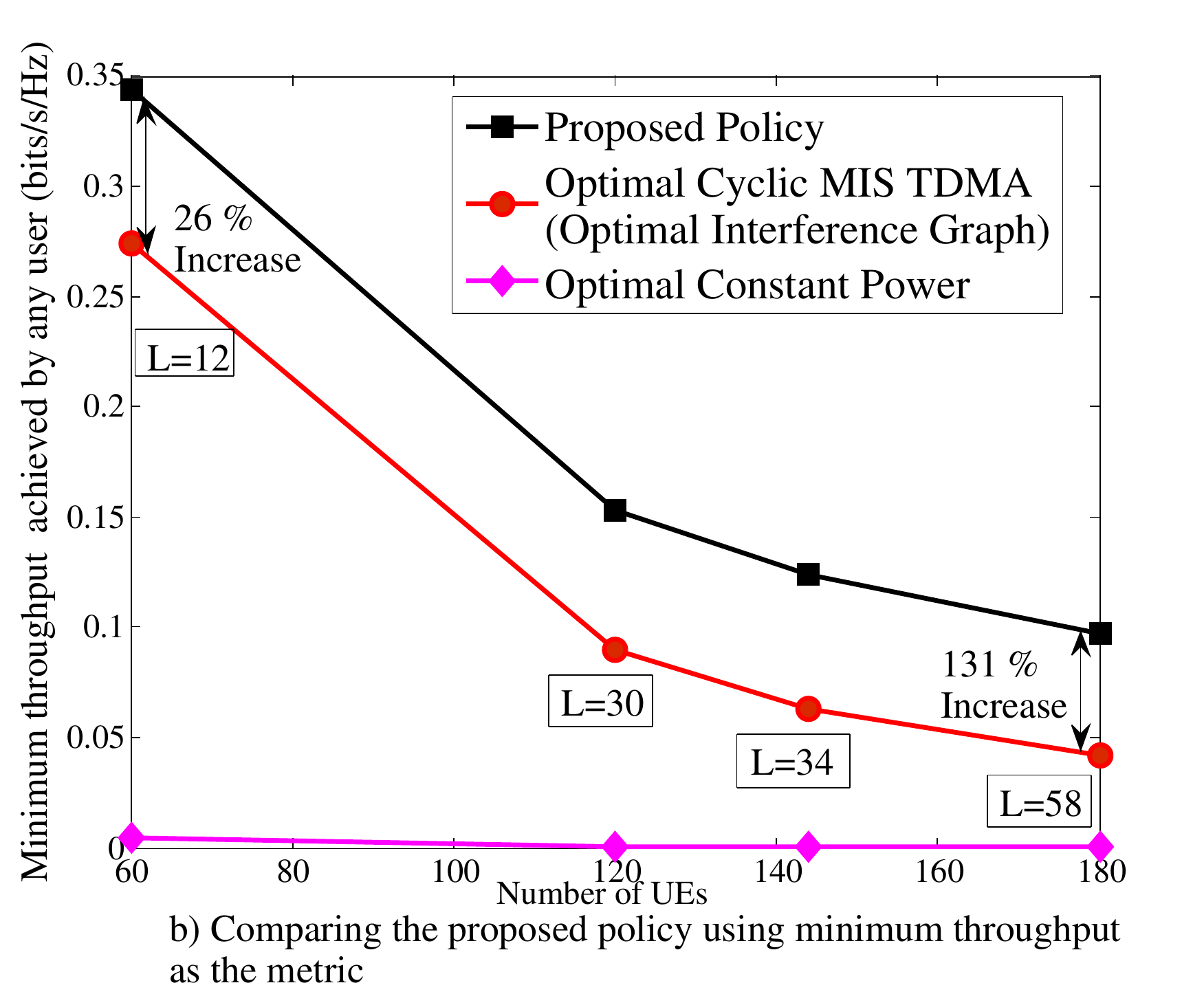}%
\end{minipage}\protect\caption{Comparing the proposed policy against others for a) sum throughput
as the metric, b) minimum throughput as the metric.}
\end{figure}

\begin{figure}
\centering \includegraphics[width=2.8in]{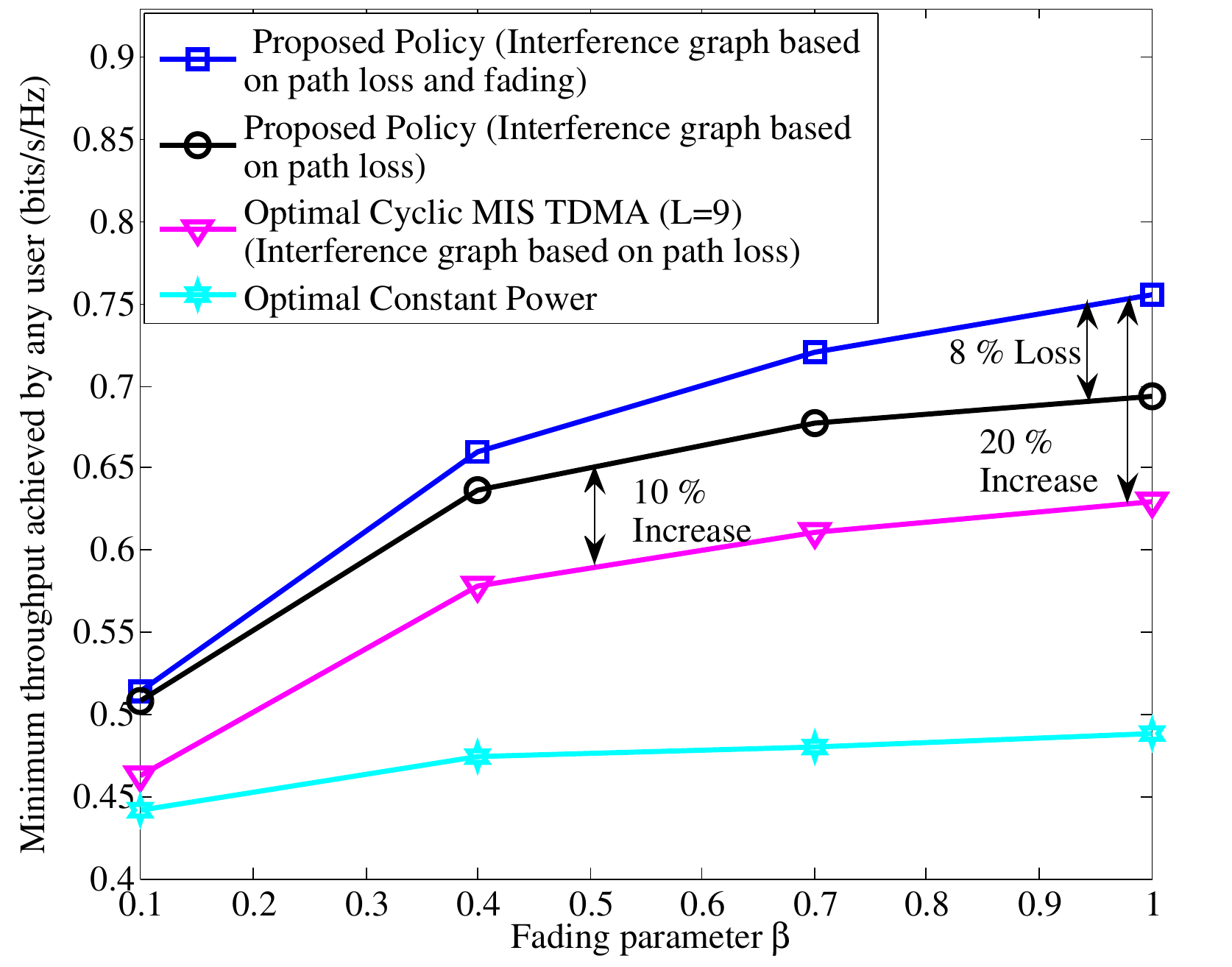}
\protect\caption{Illustrating the robustness of interference graph selection based
on path loss.}

\label{fig:9user robuse}
\end{figure}

\subsection{Performance under Dynamic Channel Conditions\label{sub:Robustness-to-Shadow}}

We consider a 9-cell network with a grid size $d=4.74$ m, where each
BS is vertically above its UE at a distance of $3.16$ m as in Subsection
\ref{sub:-Performance-gain}. Each UE has a maximum power level of
1000 mW, noise power of 1 mW and $\delta=0.89$. The channel gain
is the product of path loss as in Subsection \ref{sub:-Performance-gain}
and fading component, $f_{ij}\sim Rayleigh(\beta)$ . Here, we assume
that the fading component changes every $50$ time slots independently
and the new channel conditions are reported to the designer by each
FBS as in Step-1 in Subsection \ref{sub:Steps-in-the}. The designer
has the choice of re-computing the optimal interference graph and
thereby the optimal target every 50 time slots at the cost of a higher
complexity, or choosing a fixed optimal interference graph based on
the channel gains computed from the path loss model (which will be
graph 3 in Fig. 4 a) and selecting the optimal target every 50 time
slots based on it. In Fig. 7 we compare the loss due to choosing a
fixed interference graph with choosing the optimal interference graph
every 50 time slots. We average the performance for a duration over
a total of 10000 time slots for a fixed $\beta$. In Fig. 7, we see
that for a low $\beta$, i.e. $\beta=0.1$ which implies a lower variance
in fading the loss is only $1$\% and even when $\beta$ is large,
i.e. $\beta=1$ then as well the loss is $8$\%. We also compare with
Cyclic MIS TDMA , cycle length $L=9$ and optimal constant power policy,
the performance gain with the proposed policy using a fixed interference
graph is consistently 10 $\%$ for varying fading conditions $\beta$,
while choosing the optimal interference graph can lead to a maximum
gain of $20$\%.

\section{Conclusion}

\label{sec:conclusion}

In this work, we proposed a novel and systematic method for designing
efficient interference management policies in a network of macrocell
underlaid with femtocells. The proposed framework relies on constructing
optimal interference graphs and  optimally scheduling the MIS of
the constructed graph to maximize the network performance given the
minimum throughput requirements. Importantly, the proposed policy
is non-stationary and can address the requirements of delay sensitive
users. We prove the optimality of the proposed policies under various
deployment scenarios. The proposed policy can be implemented in a
decentralized manner with low overhead of information exchange between
BSs and UEs. For large networks, we develop a low-complexity design
framework that is provably efficient. Our proposed policies achieve
significant (up to 130 $\%$) performance improvement over existing
policies, especially for dense and large-scale deployments of femtocells.

\section{Appendix }

We would begin by defining a self-generating set similar to what is
defined in repeated game theory \cite{mailath2006repeated}. However,
our definition is less restrictive since it does not involve incentive
compatibility as required in strategic user setting in repeated games.

\textbf{Definition 1. Decomposability : } A throughput vector $\textbf{v}\in\mathbb{R}^{N+M}$
is decomposable on a set $\mathcal{W}$ with respect to a discount
factor $\delta$, if there exists a power profile $\textbf{p}\in\mathcal{P}^{MIS(G_{D})}$
and a mapping $\boldsymbol{\gamma}:\mathcal{P}^{MIS(G_{D})}\rightarrow\mathcal{W}$
such that $\forall i\;\in\{1,...N+M\}$
\begin{equation}
v_{i}=(1-\delta)r_{i}(\textbf{p})+\delta\gamma_{i}(\textbf{p})
\end{equation}
 Define $\mathcal{D}(\mathcal{W},\delta)=\{\textbf{v}:\textbf{v}\text{\;\ is pure action decomposable on\;}\mathcal{W}\}$.

\textbf{Definition 2. Self-Generation :} $\mathcal{W}$ is self-generating
with respect to discount factor $\delta$, if each throughput vector
$\textbf{v}\in\mathcal{W}$ is decomposable on $\mathcal{W}$, thus
$\mathcal{W}\subseteq\mathcal{D}(\mathcal{W},\delta)$

\textbf{Theorem 1:} Given the interference graph $G_{D}$, for any
$\delta\geq\bar{\delta}=1-\frac{1}{s(G_{D})}$, the set of throughput
profiles achieved by MIS-based policies is $\mathcal{V}^{MIS(G_{D})}(\delta)=\mathcal{V}^{MIS(G_{D})}$.

\textbf{Proof:} $\mathcal{V}^{MIS(G_{D})}=\text{conv}\{\mathcal{R}^{MIS(G_{D})}\}$,
where $\mathcal{R}^{MIS(G_{D})}=\{\textbf{r}(\textbf{p}^{I_{j}^{G_{D}}}),\forall j\in\{1,...,s(G_{D})\}\}$
and $\text{conv}\{X\}$ is the convex hull of the set $X$. We will
show that $\mathcal{V}^{MIS(G_{D})}$ is self-generating for $\delta\geq\bar{\delta}$.
If we can show that the set is self-generating then we can contruct
an MIS-based policy which achieves any given target vector in that
set. This is explained as follows. Let us assume that $\mathcal{V}^{MIS(G_{D})}$
is self-generating with respect to certain $\delta$, then given a
vector $\textbf{v}\in\mathcal{V}^{MIS(G_{D})}$ it can be decomposed
as, $\textbf{v}=(1-\delta)\textbf{r}(\textbf{p}^{I_{j}^{G_{D}}})+\delta\boldsymbol{\gamma}(\textbf{p}^{I_{j}^{G_{D}}})$.
The vector $\boldsymbol{\gamma}(\textbf{p}^{I_{j}^{G_{D}}})$ obtained
on decomposition is treated as the target vector for the transmissions
starting the next period. We know that $\boldsymbol{\gamma}(\textbf{p}^{I_{j}^{G_{D}}})\in\mathcal{V}^{MIS(G_{D})}$
and $\mathcal{V}^{MIS(G_{D})}$ is assumed to be self-generating,
which means $\boldsymbol{\gamma}(\textbf{p}^{I_{j}^{G_{D}}})$ can
also be decomposed in the same manner by a certain MIS based power
profile $\textbf{p}^{I_{j}^{G_{D}}}\in\mathcal{R}^{MIS(G_{D})}$.
This step can be recursively followed for all the future periods to
generate a policy and hence, the target $\textbf{v}$ is achieved.
Next, we show the conditions under which the set $\mathcal{V}^{MIS(G_{D})}$
is indeed self-generating.

Let $\textbf{v}\in\mathcal{V}^{MIS(G_{D})}$ and we can express $\textbf{v}=\sum_{k=1}^{s(G_{D})}\alpha_{k}\textbf{r}(\textbf{p}^{I_{k}^{G_{D}}}),\;\alpha_{k}\geq0,\;\forall k\in\{1,....,s(G_{D})\}\;\text{and }$

$\sum_{k=1}^{s(G_{D})}\alpha_{k}=1$. If $\mathcal{V}^{MIS(G_{D})}$
is self-generating with respect to certain discount factor $\delta$
then $\exists\;\textbf{p}^{I_{j}^{G_{D}}}\;\text{and}\;\boldsymbol{\gamma}(\textbf{p}^{I_{j}})\in\mathcal{V}^{MIS(G_{D})}$
about which $\textbf{v}$ can be decomposed as follows
\begin{eqnarray}
 &  & \textbf{v}=(1-\delta)\textbf{r}(\textbf{p}^{I_{j}^{G_{D}}})+\delta\boldsymbol{\gamma}(\textbf{p}^{I_{j}^{G_{D}}})
\end{eqnarray}
 Next, we come up with sufficient conditions for $\boldsymbol{\gamma}(\textbf{p}^{I_{j}^{G_{D}}})\in\mathcal{V}^{MIS(G_{D})}$
\begin{eqnarray*}
 &  & \boldsymbol{\gamma}(\textbf{p}^{I_{j}^{G_{D}}})=\frac{\textbf{v}-(1-\delta)\textbf{r}(\textbf{p}^{I_{j}^{G_{D}}})}{\delta}\\
 &  & =\sum_{k=1,k\not=j}^{s}\frac{\alpha_{k}}{\delta}\textbf{r}(\textbf{p}^{I_{k}^{G_{D}}})+\frac{\alpha_{j}-(1-\delta)}{\delta}\textbf{r}(\textbf{p}^{I_{j}^{G_{D}}})
\end{eqnarray*}

Observe that $\sum_{k=1}^{s}\frac{\alpha_{k}}{\delta}+\frac{\alpha_{j}-(1-\delta)}{\delta}=1$.
If $0\leq\frac{\alpha_{k}}{\delta}\leq1,\forall k\in\{1,...,s(G_{D})\},k\not=j$
and $0\leq\frac{\alpha_{j}-(1-\delta)}{\delta}\leq1$ then $\boldsymbol{\gamma}(\textbf{p}^{I_{j}^{G_{D}}})\in\mathcal{V}^{MIS(G_{D})}$.
This can be combined into one condition on $\delta$ given as
\[
\delta\geq\{\max_{k\not=j}\{\alpha_{k}\},1-\alpha_{j}\}
\]
Note that decomposition condition requires for the existence of atleast
one profile $\textbf{p}^{I_{j}^{G_{D}}}$. This means we can choose
the least possible bound on $\delta$, which is sufficient to ensure
that there will exist at least one profile $\textbf{p}^{I_{j}^{G_{D}}}$for
decomposition.
\begin{eqnarray*}
\delta &  & \geq\min_{j\in\{1,..s(G_{D})\}}\{\max_{k\not=j}\{\alpha_{k}\},1-\alpha_{j}\},\\
 &  & =1-\alpha_{[s(G_{D})]},
\end{eqnarray*}
 where $\alpha_{[s(G_{D})]}=\max_{j\in\{1,...,s(G_{D})\}}\{\alpha_{1},...,\alpha_{s(G_{D})}\}$.
Also, $\sum_{k=1}^{s(G_{D})}\alpha_{k}=1,\;\alpha_{k}\geq0,\;\forall k\in\{1,...,s(G_{D})\}$
yields that $\alpha_{[s(G_{D})]}\geq\frac{1}{s(G_{D})}$. Hence, the
condition $\delta\geq1-\frac{1}{s(G_{D})}$ is sufficient to ensure
decomposition of every vector in the convex hull. Thus, $\mathcal{V}^{MIS(G_{D})}$
is self-generating for all the discount factors $\delta\geq\bar{\delta}=1-\frac{1}{s(G_{D})}$.
Hence, we have been able to show that for a discount factor $\delta\geq1-\frac{1}{s(G_{D})}$
each vector in $\mathcal{V}^{MIS(G_{D})}$ can be achieved by an MIS
based policy, i.e. $\mathcal{V}^{MIS(G_{D})}\subseteq\mathcal{V}^{MIS(G_{D})}(\delta)$.
The throughput achieved by any MIS based policy is $\boldsymbol{R}(\boldsymbol{\pi})=(1-\delta)\sum_{t=0}^{\infty}\delta^{t}\textbf{r}(\boldsymbol{p}^{t})$,
here $\textbf{r}(\boldsymbol{p}^{t})\in\mathcal{R}^{MIS(G_{D})}$.
Since the coefficients $(1-\delta)\delta^{t}\geq0$ and the sum of
the coefficients of the throughput vector sum to 1, i.e. $\sum_{t=0}^{\infty}(1-\delta)\delta^{t}=1$,
this implies $\boldsymbol{R}(\boldsymbol{\pi})\in\mathcal{V}^{MIS(G_{D})}$.
Hence, $\mathcal{V}^{MIS(G_{D})}(\delta)\subseteq\mathcal{V}^{MIS(G_{D})}$.
Therefore, $\mathcal{V}^{MIS(G_{D})}(\delta)=\mathcal{V}^{MIS(G_{D})}$
for $\delta\geq1-\frac{1}{s(G_{D})}$. Next, we give a corollary of
the above Theorem 1, which states the restriction on the discount
factor and the corresponding achievable set for the policy based on
the subset of MISs in Section VI.

~~~~~~~~~~~~~~~~~~~~~~~~~~~~~~~~~~~~~~~~~~~~~~~~~~~~~~~~~~~~~~~~~~~~~~~~~~~~~~~~~~~~~~~~~~~~~~~~~~~~~~~~~~~~~~~~~~~~~~~~~~~~~~~~~(Q.E.D)

\textbf{Corollary 1:} Given the interference graph $G_{D}$, if the
$\delta\geq1-\frac{1}{C(G_{D})+1}$ then the throughput vectors achieved
by the policies in $\Pi_{approx}^{MIS(G_{D})}$ is $\mathcal{V}_{approx}^{MIS(G_{D})}$.

\textbf{Proof:} This follows on the same lines as Theorem 1. $\mathcal{V}_{approx}^{MIS(G_{D})}=\text{conv}\{\mathcal{R}_{approx}^{MIS(G_{D})}\}$,
where $\mathcal{R}_{approx}^{MIS(G_{D})}=\{\textbf{r}(\textbf{p}^{I_{j}^{G_{D}}}),\forall j\in\{k(1),...,k(C(G_{D})+1)\}\}$.
It can be shown on the same lines that $\mathcal{V}_{approx}^{MIS(G_{D})}$
is self-generating if $\delta\geq1-\frac{1}{C(G_{D})+1}$. This is
due to the fact that $\mathcal{V}_{approx}^{MIS(G_{D})}$ has $C(G_{D})+1$
extreme points.

~~~~~~~~~~~~~~~~~~~~~~~~~~~~~~~~~~~~~~~~~~~~~~~~~~~~~~~~~~~~~~~~~~~~~~~~~~~~~~~~~~~~~~~~~~~~~~~~~~~~~~~~~~~~~~~~~~~~~~~~~~~~~~~~~(Q.E.D)\vspace{1em}

\textbf{Theorem 2:} For any $\delta\geq\bar{\delta}=1-\frac{1}{s(G_{D})}$
the policy computed in Table 2 (given in Section V) achieves the target
throughput $\textbf{y}^{*}(G_{D})$.

\textbf{Proof:} The policy in Table 2 (given in the paper) is based
on the decomposition property of $\mathcal{V}^{MIS(G_{D})}$ (explained
in the proof of Theorem 1). Define $\boldsymbol{\gamma}(t)=\sum\limits _{k=1}^{s(G_{D})}\alpha_{k}(t)\textbf{r}(\textbf{p}^{I_{k}^{G_{D}}})$,
where $\alpha_{k}(t),\;\forall k\in\{1,...,s(G_{D})\},\;\forall\;t\geq0$
correspond to the coefficient $\boldsymbol{\alpha}=(\alpha_{1},...,\alpha_{N+M})$
at the beginning of time slot $t$ in the policy in Table 2. Also,
let $\textbf{p}^{I_{r_{t}}^{G_{D}}}$ correspond to the power vector
used for transmission at time $t$ for $t\geq0$. First, we show that
if $\delta\geq\bar{\delta}$ then $\boldsymbol{\gamma}(t+1)\in\mathcal{V}^{MIS(G_{D})},\;\forall t\geq0$.
Expressing $\gamma(t+1)$ in terms of coefficients of $\alpha_{k}(t),\;\forall k\in\{1,..s(G_{D})\}$
as in Table 2.
\begin{equation}
\boldsymbol{\gamma}(t+1)=\sum_{k=1}^{s(G_{D})}\alpha_{k}(t+1)\textbf{r}(\textbf{p}^{I_{k}^{G_{D}}})=\sum\limits _{k=1,k\not=r_{t}}^{s(G_{D})}\frac{\alpha_{k}(t)}{\delta}\textbf{r}(\textbf{p}^{I_{k}^{G_{D}}})+\frac{\alpha_{r_{t}}(t)-(1-\delta)}{\delta}\textbf{r}(\textbf{p}^{I_{r_{t}}^{G_{D}}})
\end{equation}
 where $\alpha_{r_{t}}(t)=\max_{k}\alpha_{k}(t)$. If $0\leq\frac{\alpha_{k}(t)}{\delta}\leq1,\forall k\in\{1,...,s(G_{D})\},k\not=r_{t}$
and $0\leq\frac{\alpha_{r_{t}}(t)-(1-\delta)}{\delta}\leq1$ then
$\boldsymbol{\gamma}(t+1)\in\mathcal{V}^{MIS(G_{D})}$. If $\delta\geq\max_{t\geq0}1-\alpha_{r_{t}}(t)$
then $\bm{\gamma}(t+1)\in\mathcal{V}^{MIS(G_{D})}\,\forall t$. We
know that $\alpha_{r_{t}}(t)\geq\frac{1}{s(G_{D})},\forall t$, this
is true because $\alpha_{r_{t}}(t)=\max_{k}\alpha_{k}(t)$. The condition
on $\delta\geq1-\frac{1}{s(G_{D})}$ implies that $\boldsymbol{\gamma}(t+1)\in\mathcal{V}^{MIS(G_{D})}$,$\forall t\geq0$.
Therefore, we can express
\[
\boldsymbol{\gamma}(t)=(1-\delta)\textbf{r}(\textbf{p}^{I_{r_{t}}^{G_{D}}})+\delta\boldsymbol{\gamma}(t+1)
\]
 here $\boldsymbol{\gamma}(t+1)\in\mathcal{V}^{MIS(G_{D})}$ . Hence
using this recursively we get,
\begin{equation}
\boldsymbol{\gamma}(0)-(1-\delta)\sum\limits _{\tau=0}^{t}\delta^{\tau}\textbf{r}(\textbf{p}^{I_{r_{\tau}}^{G_{D}}})=\delta^{t+1}\boldsymbol{\gamma}(t+1)\label{diff target}
\end{equation}
 Here, $\boldsymbol{\gamma}(t+1)\in\mathcal{V}^{MIS(G_{D})}$ and
$\boldsymbol{\gamma}(0)=\textbf{y}^{*}(G_{D})$ The above expression
in (4) specifies the difference from the target vector and the vector
achieved till time $t$. Next, we find the time $T$ after which the
norm of difference in \ref{diff target} is below $\epsilon$. $\boldsymbol{\gamma}(t+1)$
can be bounded above (since $\mathcal{V}^{MIS(G_{D})}$is closed and
bounded), $||\boldsymbol{\gamma}(t+1)||\leq\theta^{bd}$, $\forall t\geq0$.
Hence, $T=\frac{\log(\frac{\epsilon}{\theta^{bd}})}{\log(\delta)}-1$
is sufficiently high to ensure the norm of difference in \ref{diff target}
is below $\epsilon$. Hence as $\epsilon$ is chosen small, the policy
would converge to the target payoff.

Next, we give a corollary, which states the condition on the policy
based on the subset of the MISs computed in Subsection VI-A.

~~~~~~~~~~~~~~~~~~~~~~~~~~~~~~~~~~~~~~~~~~~~~~~~~~~~~~~~~~~~~~~~~~~~~~~~~~~~~~~~~~~~~~~~~~~~~~~~~~~~~~~~~~~~~~~~~~~~~~~~~~~~~~~~~(Q.E.D)

\textbf{Corollary 2:} For any $\delta\geq1-\frac{1}{C(G_{D})+1}$
the policy computed in Table 2 based on the subset of MISs computed
in Subsection VI-A achieves the target throughput $\textbf{y}_{approx}^{*}(G_{D}))$

\textbf{Proof:} The policy in Table 2 using the target weights $\boldsymbol{\alpha}_{approx}^{*}(G_{D})$
is based on the decomposition property of $\mathcal{V}_{approx}^{MIS(G_{D})}$.
Here also using the decomposition property repeatedly it can be shown
that the difference between the target and the target vector achieved
till time $t$ decreases exponentially which proves convergence.

Next, we state two lemmas which will be used to prove Theorem 3 and
4. Also, note that any inequality between two vectors would represent
a component-wise inequality.

\textbf{Lemma 1:} In a bounded degree graph with $N+M$ vertices and
bound on the maximum degree $\rho$ there exists a maximal independent
set with size $\frac{N+M}{\rho+1}$

\textbf{Proof:} In a bounded degree graph the maximum chromatic number
for the conventional vertex coloring is $\rho+1$ \cite{erciyes2013distributed}.
Let the independent sets corresponding to each color class obtained
by using the minimum coloring with atmost $\rho+1$ colors be given
as $\{I_{1}^{'},....,I_{\rho+1}^{'}\}$ and let the sets of the sizes
corresponding to each of these independent sets be given as $\{n_{1},....,n_{\rho+1}\}$.
Let the set of maximum size be $I_{[1]}^{'}$ and the corresponding
size be given as $n_{[1]}$. We know that $\sum_{i=1}^{\rho+1}n_{i}=N+M$
and $n_{[1]}\geq n_{i},\forall i\in\{1,...,\rho+1\}$. From these
conditions we can obtain $(\rho+1)n_{[1]}\geq N+M$. Hence, $n_{[1]}\geq\frac{N+M}{\rho+1}$.

~~~~~~~~~~~~~~~~~~~~~~~~~~~~~~~~~~~~~~~~~~~~~~~~~~~~~~~~~~~~~~~~~~~~~~~~~~~~~~~~~~~~~~~~~~~~~~~~~~~~~~~~~~~~~~~~~~~~~~~~~~~~~~~~~(Q.E.D)

\textbf{Lemma 2:} If the interference graph exhibits $\epsilon$-WNI
then the expression $r_{i}^{'}(\textbf{p})$ (as defined in Subsection
V-C in the paper) is an $\epsilon$ approximation to the actual rate
$r_{i}(\textbf{p})$ and this holds for all $i\in\{1,...,N+M\}$ and
$\textbf{\ensuremath{\forall}\ p}\in\mathcal{P}$.

\textbf{Proof:} To show that $r_{i}^{'}(\textbf{p})$ is an $\epsilon$
approximation, it is sufficient to show that $\max_{\textbf{p}\in\mathcal{P}}|e_{i}(\textbf{p})|\leq\epsilon$,
where $e_{i}(\textbf{p})=r_{i}^{'}(\textbf{p})-r_{i}(\textbf{p})$.
Solving for $\text{arg}\max_{\textbf{p}\in\mathcal{P}}|e_{i}(\textbf{p})|$,
we get $p_{i}=p_{i}^{max}$, $p_{j}=0,\;\forall j\;\in\mathcal{N}_{i}(G_{D})$
and $p_{k}=p_{k}^{max},\;\forall k\in(\mathcal{N}_{i}(G_{D})\cup\{i\})^{c}$.
This is explained as follows. Since the accumulative interference
from non-neighbors is not accounted for in $r_{i}^{'}(\textbf{p})$
this means $r_{i}^{'}(\textbf{p})\geq r_{i}(\textbf{p})\,\implies$
$e_{i}(\textbf{p})\geq0,\forall\textbf{p}\in\mathcal{P}$. Note that
$e_{i}(\textbf{p})=\log_{2}(\frac{1+\frac{g_{ii}p_{i}}{\sum_{j\in\mathcal{N}_{i}(G_{D})}g_{ji}p_{j}+\sigma_{i}^{2}}}{1+\frac{g_{ii}p_{i}}{\sum_{j\not=i}g_{ji}p_{j}+\sigma_{i}^{2}}})$
is an increasing function in $p_{i}$ and $p_{k},\;\forall k\in(\mathcal{N}_{i}(G_{D})\cup\{i\})^{c}$
and a decreasing function of $p_{j},\;\forall j\in\mathcal{N}_{i}(G_{D})$.
From this we get that $e_{i}(\textbf{p})$ takes its maximum value
when $p_{i}=p_{i}^{max}$ and $p_{k}=p_{k}^{max},\;\forall k\in(\mathcal{N}_{i}(G_{D})\cup\{i\})^{c}$
and $p_{j}=0,\;\forall j\in\mathcal{N}_{i}(G_{D})$. The corresponding
maximum value is $e_{i}^{max}=\log_{2}(\frac{1+\frac{g_{ii}p_{i}^{max}}{\sigma^{2}}}{1+\frac{g_{ii}p_{i}^{max}}{\sum_{\forall k\in(\mathcal{N}_{i}(G_{D})\cup\{i\})^{c}}g_{ki}p_{k}^{max}+\sigma^{2}}})$.
This combined with $\epsilon$-WNI, i.e. $Int_{i}^{max}(G_{D})=\sum_{k\in(\mathcal{N}_{i}(G_{D})\cup\{i\})^{c}}g_{ki}p_{j}^{max}\leq(2^{\epsilon}-1)\sigma_{i}^{2}$
gives that $e_{i}^{max}\leq\epsilon$. The same argument holds for
any $i$.

~~~~~~~~~~~~~~~~~~~~~~~~~~~~~~~~~~~~~~~~~~~~~~~~~~~~~~~~~~~~~~~~~~~~~~~~~~~~~~~~~~~~~~~~~~~~~~~~~~~~~~~~~~~~~~~~~~~~~~~~~~~~~~~~~(Q.E.D)

\textbf{Theorem 3:} If the constructed interference graph $G_{D^{*}}$
exhibits SLI and $\epsilon$-WNI, then the proposed policy computed
through Steps 1-5 of Section V-A achieves within $\epsilon$ of the
optimal network performance $W^{*}$.

\textbf{Proof:} With constraint tolerance of $\epsilon$, the optimization
problem in Step 5 of the design framework is stated as follows.

\begin{eqnarray}
 &  & \max_{\textbf{y},\boldsymbol{\alpha}}W(y_{1}(G_{D}),....,y_{N+M}(G_{D}))\label{des prob step 5}\\
 &  & y_{i}(G_{D})\geq R_{i}^{min}-\epsilon,\;\forall i\in\{1,...,N+M\}\nonumber \\
 &  & y_{i}(G_{D})=\sum_{k=1}^{s(G_{D})}\alpha_{k}r_{i}(\textbf{p}^{I_{k}^{G_{D}}}),\;\forall i\in\{1,...,N+M\}\nonumber \\
 &  & \alpha_{k}\geq0,\forall k\in\{1,...s(G_{D})\}\nonumber \\
 &  & \sum_{k=1}^{s(G_{D})}\alpha_{k}=1\nonumber
\end{eqnarray}

The optimal argument of the above problem is $\textbf{y}^{*}(G_{D})$.
Define $\textbf{y}^{*}(G_{D^{*}})=\text{arg}\max_{D\leq D_{max}^{BS}}W(\textbf{y}^{*}(G_{D}))$.

Consider the design problem in in Subsection IV-A (given in the paper).
The feasible region for average throughput vectors for the design
problem is a subset of $\text{conv}\{\mathcal{R}^{const}\}$. This
is explained as follows. The throughput vector achieved by an interference
management policy can be written as: $\textbf{R}(\boldsymbol{\pi})=(1-\delta)\sum_{t=0}^{\infty}\delta^{t}\textbf{r}(\textbf{p}^{t})$,
here $\textbf{R}(\boldsymbol{\pi})=(R_{1}(\boldsymbol{\pi}),...,R_{N+M}(\boldsymbol{\pi}))$.
The coefficients of $\textbf{r}(\textbf{p}^{t})$ are positive and
the sum of these coefficients is 1 and $\textbf{r}(\textbf{p}^{t})\in\mathcal{R}^{const}$,
which implies that $\textbf{R}(\boldsymbol{\pi})\in conv\{\mathcal{R}^{const}\}$.
Let $\textbf{v}^{*}\in\text{conv}\{\mathcal{R}^{const}\}$ be the
optimal solution to the design problem in Subsection IV-A (given in
the paper) with weighted sum throughput as the objective and the corresponding
optimal value of the objective is $W^{*}=\sum_{i=1}^{N+M}w_{i}v_{i}^{*}$,
where $w_{i}\geq0\;\forall i\in\{1,...,N+M\}$ and $\sum_{i=1}^{N+M}w_{i}=1$.
Note that we have assumed that the design problem in Subsection IV-A
is feasible, otherwise if it is not feasible then clearly the proposed
framework will be infeasible as well. Expressing $\textbf{v}^{*}$
in terms of throughput vectors in $\mathcal{R}^{const}$ as follows,
$\textbf{v}^{*}=\sum_{j=1}^{q}\theta_{j}\textbf{r}^{j,*}$ where $\textbf{r}^{j,*}\in\mathcal{R}^{const}$
, $\theta_{j}\geq0,\;\forall j\in\{1,...,N+M\}$ and $\sum_{j=1}^{q}\theta_{j}=1$.
Here, $\textbf{v}^{*}\geq\textbf{R}^{min}$ where $\textbf{R}^{min}=[R_{1}^{min},...,R_{N+M}^{min}]$
and the inequality between the vectors is a component-wise inequality.
Our aim is to show that there exists $\textbf{v}\in\mathcal{V}^{MIS(G_{D^{*}})}$
which is $\epsilon$ close to the optimal and satisfies the minimum
throughput constraint within a tolerance of $\epsilon$. Let $\textbf{r}(\textbf{p}^{j,*})=\textbf{r}^{j,*}$
and let $\textbf{r}^{'}(\textbf{p}^{j,*})\in\mathcal{R}_{a}^{const}$
be the corresponding throughput taking only the interference from
neighbors into account (given in Subsection V-C in the paper). Let
$\tilde{\textbf{v}}\in conv\{\mathcal{R}_{a}^{const}\}$ defined as
follows $\tilde{\textbf{v}}=\sum_{j=1}^{q}\theta_{j}\textbf{r}^{'}(\textbf{p}^{j,*})$.
Since $\textbf{r}^{'}$ is computed only from the interference contribution
of the neighbors we have $\textbf{r}^{'}(\textbf{p}^{j,*})\geq\textbf{r}(\textbf{p}^{j,*})$
and from SLI we know that there exists $\textbf{v}^{',j}\in\mathcal{V}_{a}^{MIS(G_{D^{*}})}$
which satisfies
\begin{eqnarray}
\textbf{v}^{',j}\geq\textbf{r}^{'}(\textbf{p}^{j,*})\geq\textbf{r}(\textbf{p}^{j,*}),\forall j\in\{1,....,q\}\label{ineq1 Thm 3}
\end{eqnarray}
 Let $\textbf{v}^{'}=\sum_{j=1}^{q}\theta_{j}\textbf{v}^{',j}$. Using
above \eqref{ineq1 Thm 3} we get,
\begin{equation}
\textbf{v}^{'}\geq\tilde{\textbf{v}}\geq\textbf{v}^{*}\label{ineq2 Thm 3}
\end{equation}

Expressing $\textbf{v}^{',j}$ in terms of the throughput vectors
in $\mathcal{R}_{a}^{MIS(G_{D^{*}})}$ we get $\textbf{v}^{',j}=\sum_{k=1}^{s(G_{D^{*}})}\beta_{k}^{j}\textbf{r}^{'}(\textbf{p}^{I_{k}^{G_{D^{*}}}})$
and the corresponding actual throughput is given as $\textbf{v}^{j}=\sum_{k=1}^{s(G_{D^{*}})}\beta_{k}^{j}\textbf{r}(\textbf{p}^{I_{k}^{G_{D^{*}}}})$.
From the condition in the Theorem we know $G_{D^{*}}$ exhibits $\epsilon$-WNI
which means $Int_{i}^{max}(G_{D^{*}})\leq(2^{\epsilon}-1)\sigma_{i}^{2}\;\forall i\;\in\{1,..,N+M\}$.
Hence using Lemma 2, we have $r_{i}^{'}(\textbf{p})$ is an $\epsilon$
approximation to $r_{i}(\textbf{p})$ and this holds for all $i\in\{1,...,N+M\}.$
Using $\textbf{r}(\textbf{p}^{I_{k}^{G_{D^{*}}}})\geq\textbf{r}^{'}(\textbf{p}^{I_{k}^{G_{D^{*}}}})-\epsilon,\;\forall k\in\{1,..,s(G_{D^{*}})\}$
we get
\begin{equation}
\textbf{v}^{j}\geq\textbf{v}^{',j}-\epsilon.
\end{equation}

Hence, using the lower bound on $\textbf{v}^{',j}\geq\textbf{r}(\textbf{p}^{j,*})$
we get $\textbf{v}^{j}\geq\textbf{r}(\textbf{p}^{j,*})-\epsilon$.
Also, the same can be done in general $\forall j\in\{1,...,q\}$.
Using this we get $\sum_{j=1}^{q}\theta_{j}\textbf{v}^{j}=\sum_{k=1}^{s(G_{D^{*}})}\sum_{j=1}^{q}\theta_{j}\beta_{k}^{j}\textbf{r}(\textbf{p}^{I_{k}^{G_{D^{*}}}})\geq\textbf{v}^{*}-\epsilon$.
Let $\textbf{v}=\sum_{j=1}^{q}\theta_{j}\textbf{v}^{j}$ and let $\textbf{v}=[v_{1},..,v_{N+M}]$.
We can get the following relationship.
\begin{equation}
\textbf{v}\geq\textbf{v}^{*}-\epsilon
\end{equation}
 It can be seen that we can write $\textbf{v}=\sum_{k=1}^{s(G_{D^{*}})}\alpha_{k}\textbf{r}(\textbf{p}^{I_{k}^{G_{D^{*}}}})$
with $\alpha_{k}=\sum_{j=1}^{q}\theta_{j}\beta_{k}^{j}$. Since $\alpha_{k}\geq0,\forall k\in\{1,....,s(G_{D^{*}})\}$
and $\sum_{k=1}^{s(G_{D^{*}})}\alpha_{k}=1$, which means $\textbf{v}\in\mathcal{V}^{MIS(G_{D^{*}})}$.
This gives $\sum_{i=1}^{N+M}w_{i}v_{i}\geq\sum_{i=1}^{N+M}w_{i}v_{i}^{*}-\epsilon$
and $\textbf{v}\geq\textbf{R}^{min}-\epsilon$. Hence, $\textbf{v}$
is a feasible throughput vector for \eqref{des prob step 5}. Since
$\textbf{y}^{*}(G_{D^{*}})$ is the optimal solution to the above
problem in \eqref{des prob step 5}, we can state the following,
\begin{equation}
\sum_{i=1}^{N+M}w_{i}y_{i}^{*}(G_{D^{*}})\geq\sum_{i=1}^{N+M}w_{i}v_{i}\geq\sum_{i=1}^{N+M}w_{i}v_{i}^{*}-\epsilon'\label{bound SLI}
\end{equation}
 This proves the result.

~~~~~~~~~~~~~~~~~~~~~~~~~~~~~~~~~~~~~~~~~~~~~~~~~~~~~~~~~~~~~~~~~~~~~~~~~~~~~~~~~~~~~~~~~~~~~~~~~~~~~~~~~~~~~~~~~~~~~~~~~~~~~~~~~(Q.E.D)

\textbf{Theorem 4:} For any homogeneous network with interference
graph that has a maximum degree no larger than $\rho$ and exhibits
$\zeta$-WNI with $(\rho+1)\leq\min\{\frac{\log_{2}(1+\frac{p^{max}}{\Delta^{np}2^{\zeta}\sigma^{2}})}{3R^{min}},\frac{\kappa}{(1+\rho)(\zeta)}\log_{2}(\frac{p^{max}}{\Delta^{np}\sigma^{2}})\}$,
the policy in Subsection VI-A achieves a performance $W(\textbf{y}_{approx}^{*}(G_{D}))$,
with a guarantee that $W(\textbf{y}_{approx}^{*}(G_{D}))\geq\frac{(1-\gamma)(1-\kappa)}{(1+\eta)}W(\textbf{y}^{*}(G_{D})$,
where $\gamma=(3(\rho+1))\frac{R^{min}}{\log_{2}(1+\frac{p^{max}}{\Delta^{np}2^{\epsilon}\sigma^{2}})}$.

\textbf{Proof:} We make the following homogeneity assumption $p_{i}^{max}=p^{max},R_{i}^{min}=R^{min},\max(D^{UE})_{ii}\leq\Delta,w_{i}=\frac{1}{N+M}$.
These quantities are fixed to understand the effect of scaling of
network size, i.e. $N+M$. Let $r_{i}^{max}=r_{i}(p_{i}=p^{max},\textbf{p}_{-i}=\textbf{0})$
The optimization in Step 4 in Section V-A is stated with weighted
sum throughput as the objective.
\begin{eqnarray}
\max_{\textbf{y},\boldsymbol{\alpha}} &  & \sum\limits _{i=1}^{N+M}w_{i}y_{i}\label{prb orig}\\
\text{subject to} &  & y_{i}=\sum_{j=1}^{s(G_{D})}\alpha_{j}r_{i}(\textbf{p}^{I_{j}^{G_{D}}}),\;\forall i\in\{1,...,N+M\}\nonumber \\
 &  & y_{i}\geq R^{min}\nonumber \\
 &  & \sum_{j=1}^{s(G_{D})}\alpha_{j}=1\nonumber \\
 &  & \alpha_{j}\geq0,\;\forall j\in\{1,...,s(G_{D})\}\nonumber
\end{eqnarray}

We formulate another optimization problem whose solution is an upper
bound to the above.
\begin{eqnarray}
\max_{\textbf{y},\boldsymbol{\alpha}} &  & \sum\limits _{i=1}^{N+M}w_{i}y_{i}\label{prb upbnd}\\
\text{subject to} &  & y_{i}=\sum_{j=1}^{s(G_{D})}\alpha_{j}r_{i}^{max}\textbf{1}_{i\in I_{j}^{G_{D}}},\;\forall i\in\{1,...,N+M\}\nonumber \\
 &  & \sum_{j=1}^{s(G_{D})}\alpha_{j}=1\nonumber \\
 &  & \alpha_{j}\geq0,\;\forall j\in\{1,...,s(G_{D})\}\nonumber
\end{eqnarray}

Here $\textbf{1}_{i\in I_{j}^{G_{D}}}$ is an indicator function which
takes value 1 if $i\in I_{j}^{G_{D}}$ and $0$ otherwise. Note that
the solving the above optimization problem in \eqref{prb upbnd} is
equivalent to finding the maximum weighted independent set \cite{nieberg2005robust}
with the weights assigned to each vertex $i$ given as $\bar{w}_{i}=\frac{1}{N+M}r_{i}^{max}$.
Let $\textbf{y}^{*}(G_{D})$ be the optimal solution to \eqref{prb orig}
and the corresponding optimal value is $\sum_{i=1}^{N+M}w_{i}y_{i}^{*}(G_{D})$.
Let the maximum weighted independent set which is a solution to \eqref{prb upbnd}
be denoted as $I_{p*}^{G_{D}}$ and hence the optimal value of the
objective in \eqref{prb upbnd} is $\sum_{i=1}^{N}w_{i}r_{i}^{max}\textbf{1}_{i\in I_{p*}^{G_{D}}}$.
The solution to the problem in \eqref{prb upbnd} is an upper bound
to the solution of \eqref{prb orig} , this is formally stated as
\begin{equation}
\sum_{i=1}^{N}w_{i}r_{i}^{max}\textbf{1}_{i\in I_{p*}^{G_{D}}}\geq\sum_{i=1}^{N+M}w_{i}y_{i}^{*}(G_{D})\label{eq:bound of two}
\end{equation}

This is explained as follows. Let the feasible sets of problem \eqref{prb orig}
and \eqref{prb upbnd} be given as $\mathcal{F}_{1}$, $\mathcal{F}_{2}$
respectively. If $[\boldsymbol{\alpha}^{'},\textbf{y}^{'}]\in\mathcal{F}_{1}$
then for the same $\boldsymbol{\alpha}^{'}$ the corresponding $\textbf{y}''=\sum_{j=1}^{s}\alpha_{j}^{'}\textbf{r}^{max}\textbf{1}_{i\in I_{j}^{G_{D}}}$
satisfies $\textbf{y}^{''}\geq\textbf{y}^{'}$ since $r_{i}^{max}\geq r_{i}(\textbf{p}^{I_{j}^{G_{D}}})$.

Next, we use the fact that if the interference graph exhibits $\zeta$-WNI,
i.e. $\{Int_{i}^{max}(G_{D})=\sum_{j\not\in\mathcal{N}_{i}j\not=i}g_{ji}p^{max}\leq(2^{\zeta}-1)\sigma^{2}\}$
then $r_{i}(\textbf{p})^{'}=\log_{2}(1+\frac{g_{ii}p^{max}}{\sum_{j\in\mathcal{N}_{i}(G_{D})}g_{ji}p_{j}+\sigma^{2}})$
is an $\zeta$ approximation of $r_{i}(\textbf{p})$. This follows
from Lemma 2. Thus, we have
\begin{equation}
r_{i}^{max}\textbf{1}_{i\in I_{j}^{G_{D}}}-r_{i}(\textbf{p}^{I_{j}^{G_{D}}})\leq\zeta\label{bound approx 2}
\end{equation}

The weight of approximate weighted maximum independent set $I_{k(C(G_{D})+1)}^{G_{D}}$
is given as

$\sum_{i=1}^{N+M}w_{i}r_{i}^{max}\textbf{1}_{i\in I_{k(C(G_{D})+1)}^{G_{D}}}$.
$I_{k(C(G_{D})+1)}^{G_{D}}$ is $\eta$ approximate independent set
computed using the algorithm in \cite{nieberg2005robust}, hence,
we can write
\begin{equation}
\sum_{i=1}^{N+M}w_{i}r_{i}^{max}\textbf{1}_{i\in I_{k(C(G_{D})+1)}^{G_{D}}}\geq\frac{1}{1+\eta}(\sum_{i=1}^{N+M}w_{i}r_{i}^{max}\textbf{1}_{i\in I_{p*}^{G_{D}}})\label{eta approx}
\end{equation}

If $\alpha_{k(C(G_{D})+1)}=1$ and $\alpha_{j}=0,\;\forall j\not=k(C(G_{D})+1)$
then the resulting value of the objective function is $\sum_{i=1}^{N+M}w_{i}r_{i}(\textbf{p}^{I_{k(C(G_{D})+1)}^{G_{D}}})$.
From \eqref{bound approx 2} and \eqref{eta approx} we have,
\begin{equation}
\sum_{i=1}^{N+M}w_{i}r_{i}(\textbf{p}^{I_{k(C(G_{D})+1)}^{G_{D}}})\geq\sum_{i=1}^{N+M}w_{i}r_{i}^{max}\textbf{1}_{i\in I_{k(C(G_{D})+1)}^{G_{D}}}-\zeta\geq\frac{1}{1+\eta}(\sum_{i=1}^{N+M}w_{i}r_{i}^{max}\textbf{1}_{i\in I_{p*}^{G_{D}}})-\zeta
\end{equation}

The minimum value that the expression $(\frac{1}{1+\eta}\sum_{i=1}^{N+M}w_{i}r_{i}^{max}\textbf{1}_{i\in I_{p*}^{G_{D}}})$
can take is given as $\frac{1}{(\rho+1)(\eta+1)}\log_{2}(1+\frac{1}{\Delta^{np}}\frac{p^{max}}{\sigma^{2}})$.
To derive this, first substitute the value of $w_{i}=\frac{1}{N+M}$.
From Lemma 1 we know that there exists a maximal independent set with
size no less than $\frac{N+M}{\rho+1}$. Also, using the minimum value
of the direct channel gain $g_{ii}\geq\frac{1}{\Delta^{np}}$ we get
$r_{i}^{max}\geq\log_{2}(1+\frac{1}{\Delta^{np}}\frac{p^{max}}{\sigma^{2}})$.
Combining the fact that maximal independent set has a size no less
than $\frac{N+M}{\rho+1}$ and $r_{i}^{max}\geq\log_{2}(1+\frac{1}{\Delta^{np}}\frac{p^{max}}{\sigma^{2}})$
we get the minimum value of the expression.

From the condition in the Theorem we have $\zeta<\frac{\kappa}{(\rho+1)(\eta+1)}\log_{2}(1+\frac{1}{\Delta^{np}}\frac{p^{max}}{\sigma^{2}})$,
where $0<\kappa<1$ determines the distance from the optimal solution.
If $\zeta$ is selected based on this threshold then, $\sum_{i=1}^{N+M}w_{i}r_{i}(\textbf{p}^{I_{k(C(G_{D})+1)}}\geq\frac{1-\kappa}{1+\eta}(\sum_{i=1}^{N+M}w_{i}r_{i}^{max}\textbf{1}_{v_{i}\in I_{p*}^{G_{D}}})$.
Using the fact that solution to \eqref{prb upbnd} is an upper bound
to \eqref{prb orig} as stated in \eqref{eq:bound of two}, we get
the following.
\begin{equation}
\sum_{i=1}^{N+M}w_{i}r_{i}(\textbf{p}^{I_{k(C(G_{D})+1)}})\geq\frac{1-\kappa}{1+\eta}(\sum_{i=1}^{N+M}w_{i}r_{i}^{max}\textbf{1}_{i\in I_{p*}})\geq\frac{1-\kappa}{1+\eta}\sum_{i=1}^{N+M}w_{i}y_{i}^{*}(G_{D})\label{bound approx}
\end{equation}
 Next, we state the optimization problem to compute $W(\textbf{y}_{approx}^{*}(G_{D}))$
which is similar to the problem in Step-4 in the Section V-A but uses
the subset of the MISs computed in Subsection VI-A, here $\textbf{y}_{approx}^{*}(G_{D})$
is the corresponding optimal argument. Note that $W(\textbf{y}_{approx}^{*}(G_{D}))$
is the optimal value that can be attained by the policy based on the
subset of MISs computed in Subsection VI-A in the paper.
\begin{eqnarray}
\max_{\textbf{y},\boldsymbol{\alpha}} &  & \sum\limits _{i=1}^{N+M}w_{i}y_{i}\label{prob orig approx}\\
\text{subject to} &  & y_{i}=\sum_{j=1}^{C(G_{D})+1}\alpha_{j}r_{i}(\textbf{p}^{I_{k(j)}^{G_{D}}}),\;\forall i\in\{1,...,N+M\}\nonumber \\
 &  & y_{i}\geq R^{min},\;\forall i\in\{1,...,N+M\}\nonumber \\
 &  & \sum_{j=1}^{C(G_{D})+1}\alpha_{j}=1\nonumber \\
 &  & \alpha_{j}\geq0,\;\forall j\in\{1,...,C(G_{D})+1\}\nonumber
\end{eqnarray}

Using $g_{ii}\geq\frac{1}{\Delta^{np}}$ and the fact that $Int_{i}^{max}(G_{D})\leq(2^{\zeta}-1)\sigma^{2}$
$(\zeta$-WNI) ensures that following is true
\begin{equation}
r_{i}(\textbf{p}^{I_{k(j)}^{G_{D}}})\geq\log_{2}(1+\frac{p^{max}}{\Delta^{np}2^{\zeta}\sigma^{2}})\textbf{1}_{i\in I_{k(j)}^{G_{D}}},\forall j\in\{1,...,C(G_{D})\}
\end{equation}

We now develop a feasible solution to the above problem in \eqref{prob orig approx}.
Assign $\beta_{approx,k(j)}=\frac{R^{min}}{\log_{2}(1+\frac{p^{max}}{\Delta^{np}2^{\zeta}\sigma^{2}})},\;\forall j\in\{1,...,C(G_{D})\}$.
Hence, we can write
\begin{equation}
\sum_{j=1}^{C(G_{D})}\beta_{approx,k(j)}r_{i}(\textbf{p}^{I_{k(j)}^{G_{D}}})\geq R^{min}\sum_{j=1}^{C(G_{D})}\textbf{1}_{i\in I_{k(j)}^{G_{D}}},\forall j\in\{1,...,C(G_{D})\}
\end{equation}

The maximal independent sets $\{I_{k(1)}^{G_{D}},....,I_{k(C(G_{D}))}^{G_{D}}\}$
are obtained by adding vertices in a greedy manner in ii) to the independent
sets constituting color classes, as in Subsection VI-A. Hence, all
these MISs together contain all the vertices in the graph, this ensures
$\sum_{j=1}^{C(G_{D})}\textbf{1}_{i\in I_{k(j)}^{G_{D}}}\geq1$. Therefore,
this assignment of $\beta_{approx,k(j)}$ ensures the minimum throughput
constraint in \eqref{prob orig approx} is satisfied. However, the
it still needs to be checked if $\sum_{j=1}^{C(G_{D})}\beta_{approx,k(i)}\leq1$.
Since $C(G_{D})\leq3.(\rho+1)$ from \cite{marathe1995simple}. The
condition on $\rho$, i.e. $\rho<\frac{\log_{2}(1+\frac{p^{max}}{\Delta^{np}2^{\epsilon}\sigma^{2}})}{3R^{min}}-1$
ensures $\sum_{j=1}^{C(G_{D})}\beta_{approx,k(i)}=\gamma=(3(\rho+1))\frac{R^{min}}{\log_{2}(1+\frac{p^{max}}{\Delta^{np}2^{\epsilon}\sigma^{2}})}<1$.
Hence, the solution obtained is feasible. The remaining fraction $1-\gamma$
is assigned in such a way to ensure a constant competitive ratio.

Assign $\beta_{approx,k(C(G_{D})+1)}=(1-\gamma)$ then the lower bound
is given as $W(\textbf{y}_{approx}^{*}(G_{D}))\geq(1-\gamma)\sum_{i=1}^{N+M}w_{i}r_{i}(\textbf{p}^{I_{k(C(G_{D})+1)}^{G_{D}}})$.
This combined with the lower bound in \eqref{bound approx} derived
above gives:
\begin{equation}
W(\textbf{y}_{approx}^{*}(G_{D}))\geq\frac{(1-\kappa)(1-\gamma)}{1+\eta}\sum_{i=1}^{N+M}w_{i}y_{i}^{*}(G_{D})\label{bound approx 1}
\end{equation}

~~~~~~~~~~~~~~~~~~~~~~~~~~~~~~~~~~~~~~~~~~~~~~~~~~~~~~~~~~~~~~~~~~~~~~~~~~~~~~~~~~~~~~~~~~~~~~~~~~~~~~~~~~~~~~~~~~~~~~~~~~~~~~~~~(Q.E.D)%

\subsection*{Discussion on Coloring based policies vs MIS based policies}

For this discussion we assume non-neighboring UEs in the interference
graph do not interfere at all. In a coloring based policy, the set
of UEs that are scheduled form an IS of the interference graph, which
need not be maximal. Each such IS which is not maximal can be extended
to MIS. The throughput vector obtained by scheduling UEs in a particular
IS is dominated by the throughput vector corresponding to the MIS
obtained by extending the same IS. This is because the non-neighboring
UEs which are added in order to extend IS to MIS will now have a positive
throughput and the UEs already in the IS will not be affected because
the non-neighboring UEs do not interfere.

\bibliographystyle{IEEEtran}

\bibliographystyle{IEEEtran}
\bibliography{paper_journl_jmtd_7}

\end{document}